\documentclass[10pt,twoside,english,aps,prl,superscriptaddress,twocolumn,notitlepage]{revtex4-1}
\usepackage{xcolor}
\usepackage{pdfcolmk}
\usepackage{verbatim}
\usepackage{graphicx}
\PassOptionsToPackage{normalem}{ulem}
\usepackage{ulem}
\usepackage{array}
\usepackage{verbatim}
\usepackage{textcomp}
\usepackage{multirow}

\makeatletter

\@ifundefined{textcolor}{}
{%
 \definecolor{BLACK}{gray}{0}
 \definecolor{WHITE}{gray}{1}
 \definecolor{RED}{rgb}{1,0,0}
 \definecolor{GREEN}{rgb}{0,1,0}
 \definecolor{BLUE}{rgb}{0,0,1}
 \definecolor{CYAN}{cmyk}{1,0,0,0}
 \definecolor{MAGENTA}{cmyk}{0,1,0,0}
 \definecolor{YELLOW}{cmyk}{0,0,1,0}
}

\usepackage[hypertexnames=false]{hyperref}
\hypersetup{colorlinks=true,citecolor=blue,linkcolor=black,urlcolor=blue}

\def\thesection{\Alph{section}}

\AtBeginDocument{
\renewcommand{\ref}[1]{\autoref{#1}}

}

\makeatother

\usepackage{babel}
\begin{document}

\title{Multiple Broken Symmetries in Striped La$_{2-x}$Ba$_{x}$CuO$_{4}$
detected by the \\ Field Symmetric Nernst Effect \smallskip{}}
\date{June 2015}
\author{Anjan Soumyanarayanan}\email{anjans@ntu.edu.sg}
\affiliation{Division of Physics and Applied Physics, School of Physical and Mathematical
Sciences, Nanyang Technological University, 637371 Singapore}
\affiliation{Data Storage Institute, 5 Engineering Drive 1, 117608 Singapore}
\author{X. Y. Tee}\affiliation{Division of Physics and Applied Physics, School of Physical and Mathematical
Sciences, Nanyang Technological University, 637371 Singapore}
\author{T. Ito}\affiliation{National Institute of Advanced Industrial Science and Technology,
Tsukuba, Ibaraki 305-8562, Japan}
\author{T. Ushiyama}\affiliation{National Institute of Advanced Industrial Science and Technology,
Tsukuba, Ibaraki 305-8562, Japan}
\author{Y. Tomioka}\affiliation{National Institute of Advanced Industrial Science and Technology,
Tsukuba, Ibaraki 305-8562, Japan}
\author{C. Panagopoulos}\email{christos@ntu.edu.sg}
\affiliation{Division of Physics and Applied Physics, School of Physical and Mathematical
Sciences, Nanyang Technological University, 637371 Singapore}

\begin{abstract}
We report on a thermoelectric investigation of the stripe and superconducting
phases of the cuprate La$_{2-x}$Ba$_{x}$CuO$_{4}$
near the $x=1/8$ doping known to host stable stripes. We use the
doping and magnetic field dependence of field-symmetric Nernst effect
features to delineate the phenomenology of these phases. Our measurements
are consistent with prior reports of time-reversal symmetry breaking
signatures above the superconducting $T_{{\rm c}}$, and crucially
detect a sharp, robust, field-invariant peak at the stripe charge
order temperature, $T_{{\rm {\scriptscriptstyle CO}}}$. Our observations
suggest the onset of a nontrivial charge ordered phase at $T_{{\rm {\scriptscriptstyle CO}}}$,
and the subsequent presence of spontaneously generated vortices over
a broad temperature range before the emergence of bulk superconductivity
in LBCO.
\end{abstract}
\maketitle

\section{Introduction\label{sec:IntroBG}}

\begin{comment}
\textbf{A1. Symmetry Breaking in Cuprates: }
\end{comment}
There is increasing evidence of a fundamental connection between the
phenomenology of unconventional superconductivity and the proliferation
of broken symmetry phases in hole-doped cuprates\cite{Norman2011,Lee2014,Taillefer2009}.
On one hand, numerous studies indicate the existence of electronic
phases that break translational symmetry, viz. spin/charge density
waves\cite{Sebastian2009,Wise2008} and stripes\cite{Emery1999},
and are expected to compete with or enhance superconductivity. On
the other, several reports of the onset of broken time-reversal and
point-group symmetries have recently emerged\cite{He2011c,Xia2008},
complicating the picture. A case in point is the prototypical cuprate
La$_{2-x}$Ba$_{x}$CuO$_{4}$, wherein the spin and charge of the
doped holes together form a $\sim8a_{0}$-periodic static stripe arrangement\cite{Emery1999,Fujita2004}
\textendash{} strongest at $x=1/8$ -- where bulk superconductivity
is strongly suppressed\cite{Hucker2011a}. Recent reports suggest
that striped LBCO may also host other broken symmetries, with potential
ramifications on the extensively debated superconducting mechanism
in these materials\cite{Li2011d,Karapetyan2012,Karapetyan2014,Hosur2013}.

\begin{comment}
\textbf{A2. LBCO - Recent Work: }
\end{comment}
A set of recent studies on $1/8-$LBCO have challenged the notion
that its phase diagram is well understood. First, a recent transport
study by Li \emph{et al.}\cite{Li2011d} has detected a finite Nernst
effect signal \emph{at zero magnetic field} well above $T_{{\rm c}}$,
interpreted as evidence of spontaneous vortex generation due to time-reversal
symmetry breaking (TRSB). This was corroborated by polar Kerr effect
measurements of Karapetyan \emph{et al.}\cite{Karapetyan2012}. However,
subsequent theoretical work showed that these observations could also
be consistent with a non-trivial point-group symmetry breaking (PSB)
induced by stripe charge order\cite{Hosur2013,Karapetyan2014}, wherein
the stacking of stripes in the $a-b$ plane can be modulated in a
nontrivial fashion along the $c-$axis to break inversion and mirror
symmetries. It is worth noting that the onset of stripe order plays
central role in both scenarios -- the TRSB is driven by the onset
of superconducting correlations along individual stripes, whereas
the PSB is ascribed to the long-range ordering of stripes. Several
prior observations such as that of a pairing gap and resistivity drop
above $T_{c}$\cite{Valla2006,He2008,Li2007c,Tranquada2008} support
the presence of superconducting correlations, corroborating the TRSB
picture. Meanwhile, some predictions ensuing from PSB dealing with
the varation of Kerr angle with crystal orientation have also been
verified\cite{Karapetyan2014}; however this interpretation remains
a subject of controversy\cite{Orenstein2013,Chakravarty2014,Armitage2014,Lee2014,Hosur2015}.
Importantly, neither scenario provides a fully satisfactory explanation
of recent experiments -- the sign of the TRSB signal cannot be ``trained''
by external magnetic fields, while PSB cannot explain the Nernst effect
profile, observed to peak near the onset of superconductivity\cite{Li2011d}.

\begin{comment}
\textbf{A4. Summary of Results: }
\end{comment}
Here we perform a high-resolution thermoelectric investigation of
near-$1/8$ LBCO, and use the doping and field dependence of the observed
features to delineate their behavior and understand their origin.
Our high-resolution Nernst effect measurements show TRSB signatures
consistent with prior reports\cite{Li2011d}, and further detect a
sharp, field independent peak at the stripe charge ordering temperature,
$T_{{\rm {\scriptscriptstyle CO}}}$. Our observations suggest the
onset of a nontrivial stripe charge ordered phase at $T_{{\rm {\scriptscriptstyle CO}}}$,
and the subsequent presence of spontaneously generated vortices over
a broad temperature range before the onset of bulk superconductivity
in LBCO.

\smallskip{}

\section{Methods and Results}

\begin{figure}[h]
\begin{centering}
\includegraphics[width=3.45in]{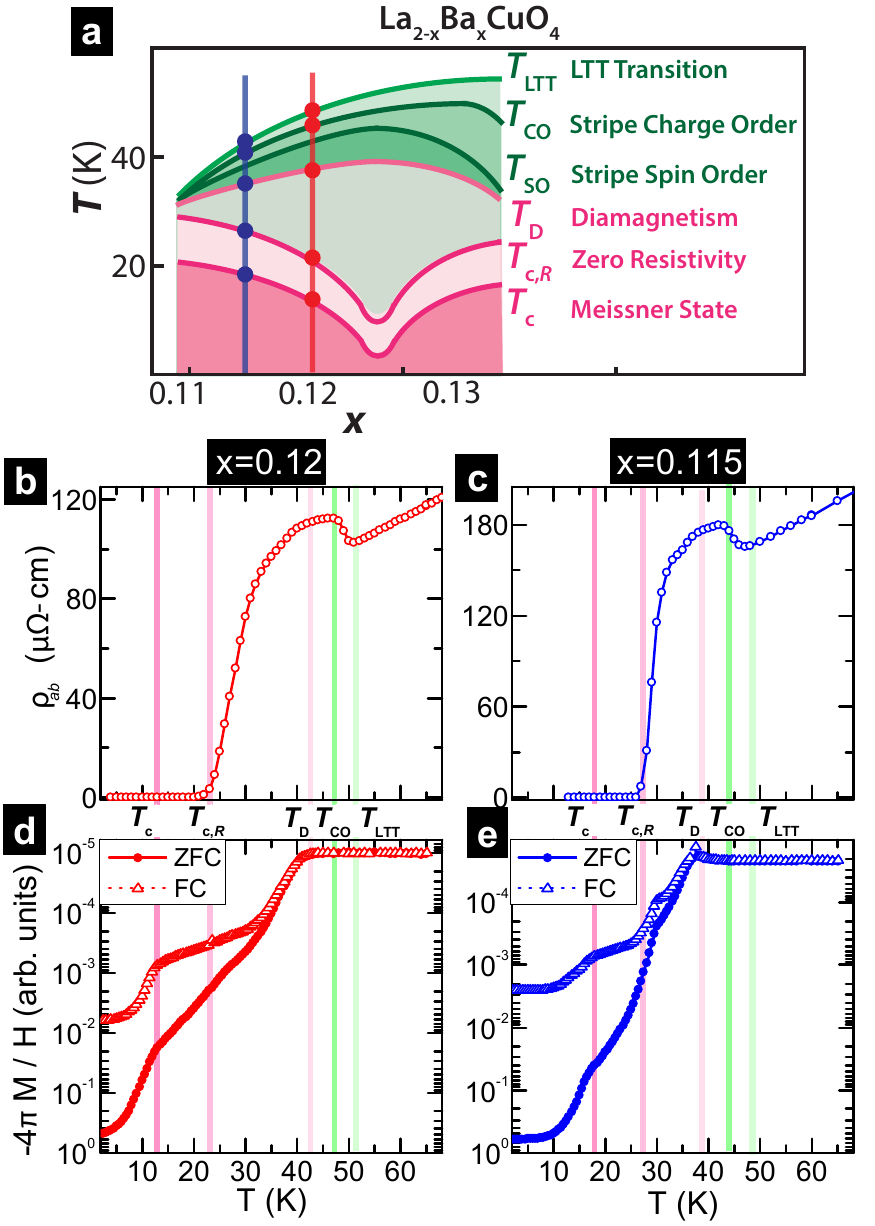}
\par\end{centering}

\protect\caption{\textbf{Transport and Magnetization. (a)} Schematic phase diagram
of La$_{2-x}$Ba$_{x}$CuO$_{4}$ around $x\sim1/8$ , with characteristic
temperature scales of stripe and superconducting phases indicated
(details in text). The relevant doping for samples studied in the
work are shown in red ($x=0.12$) and blue ($x=0.115$) respectively,
with closed circles indicating the observed transitions\cite{SOM}.
\textbf{(b-c)} In-plane resistivity ($\rho_{ab}$) and \textbf{(d-e)}
magnetization (\textbf{$H\parallel c$)} for representative samples
with $x=0.12$ (b, d) and $x=0.115$ (c, e) respectively. Temperature
scales corresponding to structural, magnetic and superconducting transitions
are identified using vertical lines, and correspond to kinks in the
resistivity and magnetization curves.\textbf{ }\label{fig:RT-MT}}
\end{figure}

\begin{comment}
\textbf{B1. Sample Prep \& Expt Setup: }
\end{comment}
{} Single crystals of La$_{2-x}$Ba$_{x}$CuO$_{4}$ (near $x=1/8$)
were grown using the recently developed laser-diode-heated floating
zone method, which enables an exceptionally high degree of sample
homogeneity\cite{Ito2013}. The samples were cut along the crystal
axes into rectangular bars for $a-b$ plane transport measurements.
For thermoelectric measurements, the typical temperature gradient
applied was $\nabla T\sim0.1$~K/mm. Further experimental methods
are detailed in \cite{SOM}.

\begin{comment}
\textbf{B2. Transport Data \& Transitions: }
\end{comment}
{} The phase diagram of hole-doped LBCO ($x\sim0.1-0.2$) has been extensively
characterized by a combination of scattering, transport and thermodynamic
measurements, establishing the signatures of structural and electronic
transitions in such measurements\cite{Li2007c,Hucker2008,Hucker2010,Hucker2011a,Li2011d}.
Our measurements of longitudinal resistivity ($\rho_{ab}$, \ref{fig:RT-MT}(b-c))
and out-of-plane magnetization ($M,\,H\parallel c$, \ref{fig:RT-MT}(d-e))
show the series of transitions that our LBCO samples undergo when
cooled below 80~K (schematic \ref{fig:RT-MT}(a)): (1) structural
transition from an orthorhombic phase to a low temperature tetragonal
(LTT) phase at $T_{{\scriptscriptstyle {\rm LTT}}}$ (48-52~K); (2)
onset of $\sim4a_{0}$ periodic charge order at $T_{{\rm {\scriptscriptstyle CO}}}$
(45-48~K); (3) onset of $\sim8a_{0}$ periodic spin order at $T_{{\rm {\scriptscriptstyle SO}}}$
($\sim$40-42~K, not detectable); (4) onset of diamagnetism at $T_{{\rm {\scriptscriptstyle D}}}$
($\sim$38-40~K); (5) zero resistivity at $T_{{\rm c},R}$ ($\sim21-26$~K);
and (6) the emergence of 3D superconductivity below $T_{{\rm c}}$
(12-18~K). Further details of the identification are discussed in
\cite{SOM}. We emphasize the quantitative consistency of these temperature
scales within our experiments\cite{SOM} and note their agreement
with existing literature\cite{Li2007c,Tranquada2008,Hucker2011a}.

\medskip{}

\noindent \textbf{\emph{Zero Field Nernst Effect}}

\begin{figure}[h]
\begin{centering}
\includegraphics[width=3.45in]{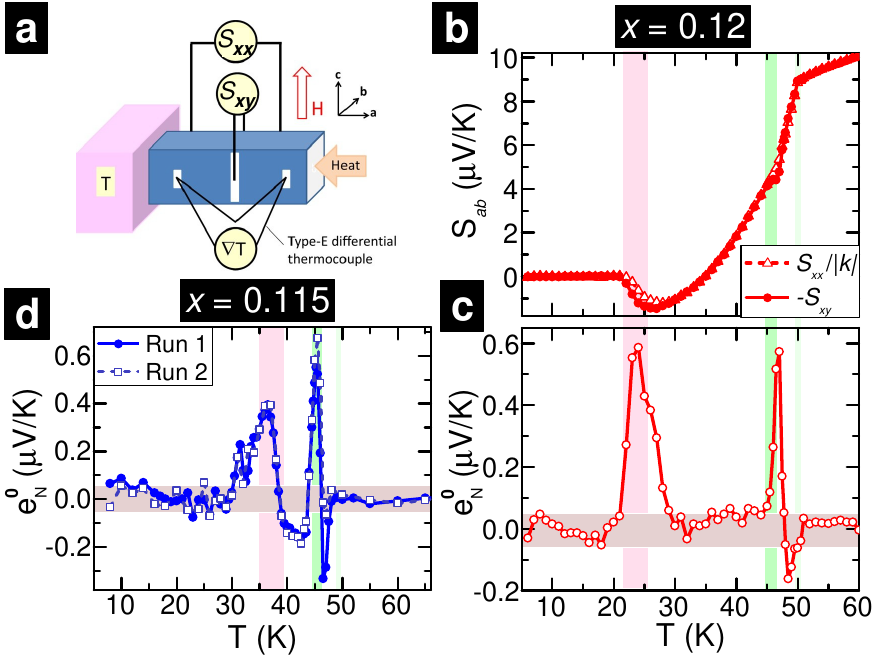}
\par\end{centering}

\protect\caption{\textbf{Zero Field Nernst Effect. (a)} Schematic contact configurations
for simultaneous Seebeck (longitudinal, or $S_{xx}$) and Nernst (transverse,
or $S_{xy}$) effect measurements. Inset shows the crystalline axes
orientation for the samples. \textbf{(b)} The measured zero field
Seebeck ($S_{xx}/|k|$, $k=-4.5$) and Nernst ($S_{xy}$) coefficients
for $x=0.12$ as a function of temperature, overlaid by superposing
their values above 55~K\cite{Li2011d}. \textbf{(c)} True zero field
Nernst (ZFN) signal, $e_{{\rm {\scriptscriptstyle N}}}^{0}(T)$ for
$x=0.12$ extracted by subtracting the longitudinal pickup ($S_{xx}(T)$)
from the measured Nernst response ($S_{xy}(T)$) using (b). Finite
contributions to $e_{{\rm {\scriptscriptstyle N}}}^{0}(T)$ are observed
just above $T_{{\rm c},R}$ and at $T_{{\scriptscriptstyle {\rm CO}}}$.
\textbf{(d)} Similarly obtained ZFN signal $e_{{\rm {\scriptscriptstyle N}}}^{0}(T)$
for $x=0.115$, showing features consistent with (c). \label{fig:ZFN_Summary}}
\end{figure}

\begin{comment}
\textbf{B3. Transport Data \& Transitions: }
\end{comment}
To distinguish the evidence for TRSB due to phase incoherent superconductivity\cite{Li2011d}
from the signatures of nontrivial charge ordering\cite{Karapetyan2014},
it is imperative to examine their evolution with doping and magnetic
field using techniques sensitive to both phenomena\cite{Hosur2013}.
The large resistivity anisotropy in LBCO ($\rho_{c}/\rho_{ab}\sim10^{3}$,
\cite{Li2007c}) severely limits attempts to detect their presence
using anomalous Hall effect -- due to unavoidable artifacts resulting
from $c$-axis pickup\cite{SOM}. In contrast, the near-isotropic
thermoelectric properties of LBCO\cite{Nakamura1993} enable the field-symmetric
Nernst coefficient, $e_{{\rm {\scriptscriptstyle N}}}^{S}(T)$, to
probe their signatures with the requisite sensitivity. Having determined
the characteristic temperature scales of stripe and superconducting
phases for our samples, we thus turn to the Nernst effect measurements
forming the nucleus of this work.

\begin{comment}
\textbf{B4. ZFN - Technique:}
\end{comment}
{} The Nernst coefficient, $e_{{\rm {\scriptscriptstyle N}}}=V_{y}/\nabla T_{x}$
corresponds to the transverse voltage $V_{y}$ generated due to a
longitudinal thermal gradient $\nabla T_{x}$. It is typically observed
at a finite magnetic field, and in the cuprates, has been attributed
to moving vortices\cite{Wang2006,Li2007a}, Gaussian fluctuations\cite{Ussishkin2002},
or to quasiparticles arising from fluctuating stripes\cite{Cyr-Choiniere2009,Chang2010}.
While the aforementioned \emph{field anti-symmetric}, or conventional
Nernst coefficient is determined entirely from transverse thermoelectric
measurements at fields $\pm H$ ($e_{{\rm {\scriptscriptstyle N}}}^{{\rm A}}=(S_{xy}(H)-S_{xy}(-H))/2$),
this is not possible for the the \emph{field-symmetric}, unconventional
component of interest to us ($e_{{\rm {\scriptscriptstyle N}}}^{{\rm {\scriptscriptstyle S}}}=(S_{xy}(H)+S_{xy}(-H))/2$).
For example, at zero field, the observed signal ($S_{xy}$) unavoidably
contains a longitudinal $S_{xx}$ pickup due to a slight misalignment
of the contact leads (\ref{fig:ZFN_Summary}b). Therefore, we obtain
the true zero field Nernst (ZFN) coefficient $e_{{\rm {\scriptscriptstyle N}}}^{0}(T)$
(and, by extension, $e_{{\rm {\scriptscriptstyle N}}}^{{\rm {\scriptscriptstyle S}}}(T)$
at finite fields) by removing the $S_{xx}$ contribution to $S_{xy}$,
i.e. $e_{{\rm {\scriptscriptstyle N}}}^{0}(T)=S_{xy}(T)-S_{xx}(T)/k$\cite{Li2011d},
where $S_{xx}(T)$ and $S_{xy}(T)$ are measured simultaneously (schematic
in \ref{fig:ZFN_Summary}a, further details in \cite{SOM}).

\begin{comment}
\textbf{B5. ZFN - Observations:}
\end{comment}
{} The ZFN coefficient $e_{{\rm {\scriptscriptstyle N}}}^{0}(T)$ measured
as a function of temperature using this compensated technique for
$x=0.12$ and $x=0.115$ respectively (\ref{fig:ZFN_Summary}c-d)
shows several features at characteristic transition temperatures that
are consistent across doping. First, $e_{{\rm {\scriptscriptstyle N}}}^{0}(T)$
is finite only for $T_{{\rm c},R}<T<T_{{\scriptscriptstyle {\rm LTT}}}$,
i.e. in the presence of static stripes, yet in the absence of bulk
superconductivity, as reported for $x=0.125$ \cite{Li2011d}. Second,
$e_{{\rm {\scriptscriptstyle N}}}^{0}(T)$ can be bipolar (\ref{fig:ZFN_Summary}d)
in contrast with \cite{Li2011d}, and the exact behavior is reproduced
through multiple temperature cycles. Third, we observe a broad hump
(width $\sim8$~K) just above $T_{{\rm c},R}$ - similar to that
reported in \cite{Li2011d}, which has been ascribed to spontaneous
vortex generation. Fourth, and crucially, we observe a \textbf{\emph{sharp
peak }}(width $\sim1$~K) at a temperature previously identified
as $T_{{\rm {\scriptscriptstyle CO}}}$. This $T_{{\rm {\scriptscriptstyle CO}}}$
peak, also visible in the raw data (\ref{fig:ZFN_Summary}b) has been
hitherto unobserved likely due to its sharp linewidth; our high temperature
resolution ($\sim0.25$~K) and small temperature gradients ($\sim0.1$~K/mm)
enable its detection. We reiterate that such a sharp peak is in contrast
to a broad hump expected from the presence of vortices\cite{Li2011d}.
Importantly, its coincidence with the onset of stripe charge order
( $T_{{\rm {\scriptscriptstyle CO}}}$) is suggestive of its origin.

\medskip{}

\noindent \textbf{\emph{Symmetric Nernst Effect: Field Dependence}}

\begin{figure}[h]
\begin{centering}
\includegraphics[width=2in]{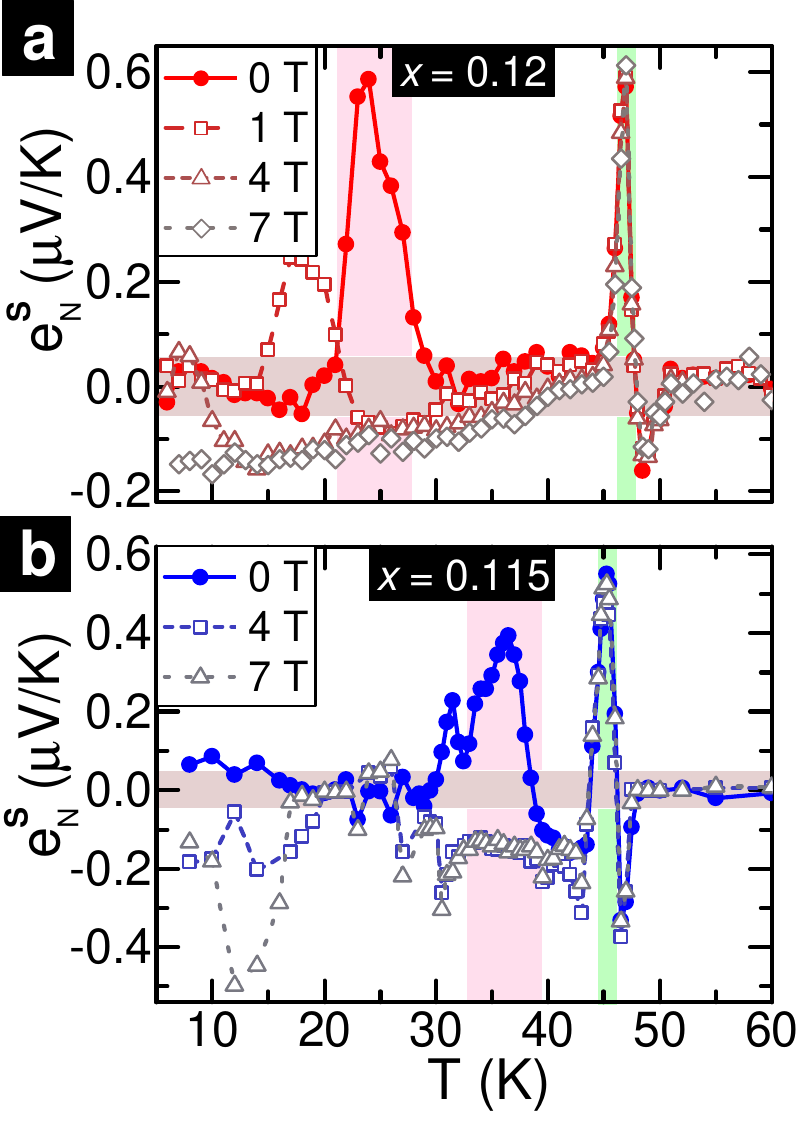}
\par\end{centering}

\protect\caption{\textbf{Field Dependence of Symmetric Nernst Effect.} The field-symmetric
part of the Nernst coefficient, $e_{{\rm {\scriptscriptstyle N}}}^{S}(T)$
at various magnetic fields for \textbf{(a)} $x=0.12$ and \textbf{(b)}
$x=0.115$. The sharp $\sim T_{{\scriptscriptstyle {\rm CO}}}$ feature
is field-invariant, while the lower temperature feature is strongly
suppressed by magnetic field. \label{fig:BNernst}}
\end{figure}

\textcolor{blue}{}%
\begin{comment}
\textbf{B6. ZFN Field-Dependence}
\end{comment}
Having identified features of interest in the ZFN data, we now turn
to the field-symmetric evolution of these data ($e_{{\rm {\scriptscriptstyle N}}}^{S}$,
shown in \ref{fig:BNernst}), to further understand the origin of
these features. The two peaked features identified previously have
remarkably contrasting field dependent behavior -- consistent across
doping. First, the broad hump just above $T_{{\rm c},R}$ is strongly
suppressed with field -- it is much reduced in magnitude and is observed
at lower temperatures. This is consistent with its expected origin
from the spontaneous generation of vortices of one sign, which would
be stabilized by TRSB\textcolor{blue}{\cite{Li2011d}. }In contrast,
the sharp peak at $T_{{\rm {\scriptscriptstyle CO}}}$ has no observable
field dependence in position or magnitude, maintaining a robust presence
at $T_{{\rm {\scriptscriptstyle CO}}}$ across field and doping. This
strongly suggests that the latter peak does not have a superconducting
origin, and could instead emerge from other nontrivial symmetry breaking
phenomena\cite{Karapetyan2014}. Finally, we also note the field
dependence of the $e_{{\rm {\scriptscriptstyle N}}}^{S}(T)$ background
emerging at or just below $T_{{\scriptscriptstyle {\rm CO}}}$, and
persisting to lower temperatures.

\medskip{}

\noindent \textbf{\emph{Conventional Nernst Effect}}

\begin{figure}[h]
\begin{centering}
\includegraphics[width=3.45in]{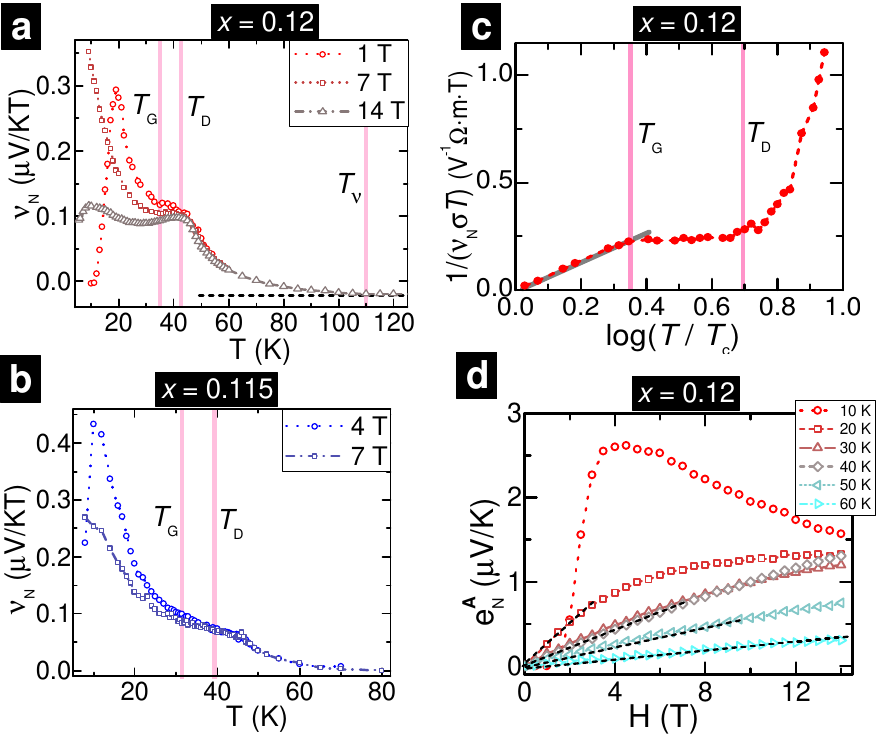}
\par\end{centering}

\protect\caption{\textbf{Conventional Nernst Effect. (a, b)} Temperature dependence
of the normalized anti-symmetric Nernst coefficient, $v_{{\rm {\scriptscriptstyle N}}}(T)=e_{{\rm {\scriptscriptstyle N}}}^{{\rm A}}(T)/H$,
at various magnetic fields for \textbf{(a)} $x=0.12$, and \textbf{(b)}
$x=0.115$. Shaded lines indicate a deviation from the constant background
($T_{{\rm \nu}}\sim110$ K), and characteristic features at $T_{{\rm {\scriptscriptstyle D}}}$
and $T_{{\rm {\rm {\scriptscriptstyle G}}}}$.\textbf{ (c)} Plot of
$1/(v_{{\rm {\scriptscriptstyle N}}}\sigma T)$ versus $\ln(T/T_{{\rm c}})$
for $x=0.12$, with the solid gray line indicating a linear fit to
Gaussian fluctuation theory\cite{Serbyn2009,Michaeli2009} below $T_{{\rm {\scriptscriptstyle G}}}=33$
K. \textbf{(d)} Magnetic field dependence of the anti-symmetric Nernst
coefficient $e_{{\scriptscriptstyle N}}^{{\rm {\scriptscriptstyle A}}}(H)$
at various temperatures for $x=0.12$, showing the onset of non-linearity
below $T_{{\rm {\scriptscriptstyle D}}}$. \label{fig:Nernst_AS}}

\end{figure}

\begin{comment}
\textbf{B7. Conventional Nernst Effect: }
\end{comment}
{} The conventional Nernst effect has been extensively utilized to probe
superconducting and quasiparticle fluctuations in the cuprates\cite{Wang2006,Cyr-Choiniere2009}.
\ref{fig:Nernst_AS}a-b show the temperature dependence of the field-normalized,
anti-symmetric Nernst coefficient of LBCO-0.120, $v_{{\scriptscriptstyle {\rm N}}}(T)=e_{{\rm {\scriptscriptstyle N}}}^{{\rm A}}(T)/H$,
measured at various magnetic fields for the two doping values. At
high temperatures, the near-constant $v_{{\rm {\scriptscriptstyle N}}}$
results from quasiparticle transport\cite{Wang2006}. Below the Nernst
onset temperature $T_{v}\sim110$ K, $v_{{\scriptscriptstyle N}}(T)$
deviates from the background, with a kink observed at $T_{{\rm {\scriptscriptstyle CO}}}$.
The suppression of $v_{{\rm {\scriptscriptstyle N}}}$ by magnetic
field at temperatures below the onset of diamagnetism, $T_{{\rm {\scriptscriptstyle D}}}$,
is a clear signature of superconducting vortices \cite{Xu2000,Wang2006,Ussishkin2002,Kokanovic2009}.
This may result from either vortex excitations produced by phase fluctuations\cite{Xu2000,Wang2006}
or from Gaussian amplitude fluctuations caused by short-lived Cooper
pairs\cite{Ussishkin2002,Kokanovic2009}. In our case, $v_{{\rm {\scriptscriptstyle N}}}$
increases dramatically below $T_{{\rm {\scriptscriptstyle G}}}=33$
K for $x=0.12$ -- the onset temperature of Gaussian fluctuations.
This is verified in \ref{fig:Nernst_AS}c, wherein a linear relationship
is observed between $1/(v_{{\scriptscriptstyle {\rm N}}}\sigma T)$
and $\ln(T/T_{{\rm c}})$ for $T_{{\rm {\rm c}}}<T<T_{{\rm {\scriptscriptstyle G}}}$
-- indicating the dominance of Gaussian fluctuations\cite{Ussishkin2002,Serbyn2009,Michaeli2009}.
Meanwhile, the superconducting signatures observed over $T_{{\rm {\scriptscriptstyle G}}}<T<T_{{\rm {\scriptscriptstyle D}}}$
are likely due to vortex excitations. We note that the broad ZFN peak
occurs in this temperature regime, consistent with its ascribed origin
to superconducting vortices resulting from TRSB\cite{Li2011d}. Finally,
the absence of any measurable field-dependence signatures in $v_{{\rm {\scriptscriptstyle N}}}$
at $T_{{\rm {\scriptscriptstyle CO}}}$ corroborate the non-superconducting
origin of the $T_{{\rm {\scriptscriptstyle CO}}}$ peak in the complementary
ZFN measurements.

\section{Discussion}

\begin{comment}
\textbf{C1. ZFN Peak Origin - Extrinsic Effects:}
\end{comment}
Our detailed doping and field-dependent studies show that the symmetric
Nernst effect signal observed in near-$1/8$ LBCO is comprised of
two distinct components -- a broad, field-dependent hump above $T_{{\rm c},R}$
of superconducting origin, and a sharp field-independent peak at $T_{{\rm {\scriptscriptstyle CO}}}$.
The observed behavior of the broad hump with doping and field substantiates
prior reports at $x=1/8$ of spontaneously generated vortices arising
from TRSB\cite{Li2011d}. In contrast, the hitherto unobserved robust
peak at $T_{{\rm {\scriptscriptstyle CO}}}$, unchanged with field,
is suggestive of a stripe charge order origin. It is worth emphasizing
that the observed result is a large, spontaneously generated\emph{
transverse electric field} at $T_{{\rm {\scriptscriptstyle CO}}}$
in response to a longitudinal thermal gradient. Having ruled out a
superconducting origin, we consider the possibility that this results
from the heat current going off-axis for extrinsic reasons, either
induced by the contacts, or sample inhomogeneity. First, we point
out that repeated (3-5 times) measurements across multiple samples
for each doping with fresh electrical and thermal contacts show a
$T_{{\rm {\scriptscriptstyle CO}}}$ peak constant in magnitude within
measurement error -- discounting contact-related artifacts.\textcolor{blue}{{}
}Next, the possibility of marked physical or chemical inhomogeneity
deeper inside the sample causing this effect can also be ruled out
as these should be detectable within the complementary $\rho_{xx}(T)$,
$\rho_{xy}(T)$, $S_{xx}(T)$ and $M(T)$ measurements, which are
instead consistent with the expected behavior at the corresponding
doping\cite{SOM}. Finally, in a perfectly crystalline sample, is
also possible that the onset of unidirectional charge stripes drastically
alters the thermal transport properties, introducing a transverse
component to the thermal current. Here it is worth noting that our
samples are not detwinned, and such unidirectional behavior is expected
to average out over the sample size, as evidenced in Hall measurements\cite{SOM}.
Moreover, our measurements of transverse thermal gradients across
this temperature range show that any such transverse effects would
be an order of magnitude smaller than the the observed sharp, sizeable
feature at $T_{{\rm {\scriptscriptstyle CO}}}$\cite{SOM}. Furthermore,
since the sample is heated to high-temperatures ($\sim$ 700~K, for
contact preparation) between successive measurements, the presence
of a quantitatively reproducible peak is not tenable in such a scenario.

\begin{comment}
\textbf{C2. ZFN Peak Origin - Intrinsic Effects:}
\end{comment}
Finally, we examine plausible scenarios wherein the $T_{{\rm {\scriptscriptstyle CO}}}$
peak emerges from intrinsic effects resulting from the onset of stripe
charge order. One possibility is that the presence of some tetragonal
symmetry breaking could result in the mixing of longitudinal and transverse
transport coefficients. While such effects would be small or absent
in a perfectly tetragonal crystal, they could be induced by charge
or superconducting stripes, and would be preferentially oriented along
crystallographic directions. Another possibility is that this peak
results from the point-group symmetry breaking emerging from nontrivial
stacking of stripes\cite{Hosur2013,Hosur2015}, as observed in Kerr
effect measurements of similar samples\cite{Karapetyan2012}. In this
latter case, one would expect PSB, and thus the ZFN signal to persist
to well below $T_{{\rm {\scriptscriptstyle CO}}}$. While the peak-like
manifestation of the ZFN signal could, in principle, result from the
interplay of a PSB signal and the field-dependent background, a quantitative
explanation of the observations is imperative.

\begin{comment}
\textbf{C3. Conclusions:}
\end{comment}
In summary, we have performed a detailed investigation of the thermoelectric
coefficients of near-$1/8$ LBCO, with varying doping and magnetic
field. Our symmetric Nernst effect signal is comprised of two distinct
components -- a broad, field-dependent hump above $T_{{\rm c},R}$
of superconducting origin, and a sharp field-independent peak at $T_{{\rm {\scriptscriptstyle CO}}}$.
While the former is consistent with prior reports indicative of spontaneous
TRSB, the latter, likely of stripe charge order origin, merits a comprehensive
theoretical explanation.

\noindent \vspace{0.5ex}

\noindent We are grateful to Ivar Martin for insightful discussions.
This work was supported by the National Research Foundation, Singapore,
through Grant NRF-CRP4-2008-04. The work at AIST was supported by
JSPS Grants-in-Aid for Scientific Research (Grant no. 22560018). A.S.
and X.Y.T. contributed equally to this work.

%%%%%%%%%%%%%%%%%%%%%%%%%%%%%%%%%%%%%%%%%%%%%%%%%%%%%%%%%%%%%%%%%%%%%%
%%%%%%%%%%%%%%%%%%%%%%%%%%%%%%%%%%%%%%%%%%%%%%%%%%%%%%%%%%%%%%%%%%%%%%
%%%%% Bibliography
%%%%%%%%%%%%%%%%%%%%%%%%%%%%%%%%%%%%%%%%%%%%%%%%%%%%%%%%%%%%%%%%%%%%%%
%%%%%%%%%%%%%%%%%%%%%%%%%%%%%%%%%%%%%%%%%%%%%%%%%%%%%%%%%%%%%%%%%%%%%%
\noindent \bibliographystyle{apsrev4-1}

\begin{thebibliography}{36}%
\makeatletter
\providecommand \@ifxundefined [1]{%
 \@ifx{#1\undefined}
}%
\providecommand \@ifnum [1]{%
 \ifnum #1\expandafter \@firstoftwo
 \else \expandafter \@secondoftwo
 \fi
}%
\providecommand \@ifx [1]{%
 \ifx #1\expandafter \@firstoftwo
 \else \expandafter \@secondoftwo
 \fi
}%
\providecommand \natexlab [1]{#1}%
\providecommand \enquote  [1]{``#1''}%
\providecommand \bibnamefont  [1]{#1}%
\providecommand \bibfnamefont [1]{#1}%
\providecommand \citenamefont [1]{#1}%
\providecommand \href@noop [0]{\@secondoftwo}%
\providecommand \href [0]{\begingroup \@sanitize@url \@href}%
\providecommand \@href[1]{\@@startlink{#1}\@@href}%
\providecommand \@@href[1]{\endgroup#1\@@endlink}%
\providecommand \@sanitize@url [0]{\catcode `\\12\catcode `\$12\catcode
  `\&12\catcode `\#12\catcode `\^12\catcode `\_12\catcode `\%12\relax}%
\providecommand \@@startlink[1]{}%
\providecommand \@@endlink[0]{}%
\providecommand \url  [0]{\begingroup\@sanitize@url \@url }%
\providecommand \@url [1]{\endgroup\@href {#1}{\urlprefix }}%
\providecommand \urlprefix  [0]{URL }%
\providecommand \Eprint [0]{\href }%
\providecommand \doibase [0]{http://dx.doi.org/}%
\providecommand \selectlanguage [0]{\@gobble}%
\providecommand \bibinfo  [0]{\@secondoftwo}%
\providecommand \bibfield  [0]{\@secondoftwo}%
\providecommand \translation [1]{[#1]}%
\providecommand \BibitemOpen [0]{}%
\providecommand \bibitemStop [0]{}%
\providecommand \bibitemNoStop [0]{.\EOS\space}%
\providecommand \EOS [0]{\spacefactor3000\relax}%
\providecommand \BibitemShut  [1]{\csname bibitem#1\endcsname}%
\let\auto@bib@innerbib\@empty
%</preamble>
\bibitem [{\citenamefont {Norman}(2011)}]{Norman2011}%
  \BibitemOpen
  \bibfield  {author} {\bibinfo {author} {\bibfnamefont {M.~R.}\ \bibnamefont
  {Norman}},\ }\href {\doibase 10.1126/science.1200181} {\bibfield  {journal}
  {\bibinfo  {journal} {Science}\ }\textbf {\bibinfo {volume} {332}},\ \bibinfo
  {pages} {196} (\bibinfo {year} {2011})}\BibitemShut {NoStop}%
\bibitem [{\citenamefont {Lee}(2014)}]{Lee2014}%
  \BibitemOpen
  \bibfield  {author} {\bibinfo {author} {\bibfnamefont {P.~A.}\ \bibnamefont
  {Lee}},\ }\href {\doibase 10.1103/PhysRevX.4.031017} {\bibfield  {journal}
  {\bibinfo  {journal} {Phys. Rev. X}\ }\textbf {\bibinfo {volume} {4}},\
  \bibinfo {pages} {031017} (\bibinfo {year} {2014})}\BibitemShut {NoStop}%
\bibitem [{\citenamefont {Taillefer}(2009)}]{Taillefer2009}%
  \BibitemOpen
  \bibfield  {author} {\bibinfo {author} {\bibfnamefont {L.}~\bibnamefont
  {Taillefer}},\ }\href {\doibase 10.1088/0953-8984/21/16/164212} {\bibfield
  {journal} {\bibinfo  {journal} {J. Phys.: Condens. Matter}\ }\textbf
  {\bibinfo {volume} {21}},\ \bibinfo {pages} {164212} (\bibinfo {year}
  {2009})}\BibitemShut {NoStop}%
\bibitem [{\citenamefont {Sebastian}\ \emph {et~al.}(2009)\citenamefont
  {Sebastian}, \citenamefont {Harrison}, \citenamefont {Mielke}, \citenamefont
  {Liang}, \citenamefont {Bonn}, \citenamefont {Hardy},\ and\ \citenamefont
  {Lonzarich}}]{Sebastian2009}%
  \BibitemOpen
  \bibfield  {author} {\bibinfo {author} {\bibfnamefont {S.~E.}\ \bibnamefont
  {Sebastian}}, \bibinfo {author} {\bibfnamefont {N.}~\bibnamefont {Harrison}},
  \bibinfo {author} {\bibfnamefont {C.~H.}\ \bibnamefont {Mielke}}, \bibinfo
  {author} {\bibfnamefont {R.}~\bibnamefont {Liang}}, \bibinfo {author}
  {\bibfnamefont {D.~A.}\ \bibnamefont {Bonn}}, \bibinfo {author}
  {\bibfnamefont {W.~N.}\ \bibnamefont {Hardy}}, \ and\ \bibinfo {author}
  {\bibfnamefont {G.~G.}\ \bibnamefont {Lonzarich}},\ }\href {\doibase
  10.1103/PhysRevLett.103.256405} {\bibfield  {journal} {\bibinfo  {journal}
  {Phys. Rev. Lett.}\ }\textbf {\bibinfo {volume} {103}},\ \bibinfo {pages} {1}
  (\bibinfo {year} {2009})}\BibitemShut {NoStop}%
\bibitem [{\citenamefont {Wise}\ \emph {et~al.}(2008)\citenamefont {Wise},
  \citenamefont {Boyer}, \citenamefont {Chatterjee}, \citenamefont {Kondo},
  \citenamefont {Takeuchi}, \citenamefont {Ikuta}, \citenamefont {Wang},\ and\
  \citenamefont {Hudson}}]{Wise2008}%
  \BibitemOpen
  \bibfield  {author} {\bibinfo {author} {\bibfnamefont {W.~D.}\ \bibnamefont
  {Wise}}, \bibinfo {author} {\bibfnamefont {M.~C.}\ \bibnamefont {Boyer}},
  \bibinfo {author} {\bibfnamefont {K.}~\bibnamefont {Chatterjee}}, \bibinfo
  {author} {\bibfnamefont {T.}~\bibnamefont {Kondo}}, \bibinfo {author}
  {\bibfnamefont {T.}~\bibnamefont {Takeuchi}}, \bibinfo {author}
  {\bibfnamefont {H.}~\bibnamefont {Ikuta}}, \bibinfo {author} {\bibfnamefont
  {Y.}~\bibnamefont {Wang}}, \ and\ \bibinfo {author} {\bibfnamefont {E.~W.}\
  \bibnamefont {Hudson}},\ }\href {\doibase 10.1038/nphys1021} {\bibfield
  {journal} {\bibinfo  {journal} {Nat. Phys.}\ }\textbf {\bibinfo {volume}
  {4}},\ \bibinfo {pages} {696} (\bibinfo {year} {2008})}\BibitemShut {NoStop}%
\bibitem [{\citenamefont {Emery}\ \emph {et~al.}(1999)\citenamefont {Emery},
  \citenamefont {Kivelson},\ and\ \citenamefont {Tranquada}}]{Emery1999}%
  \BibitemOpen
  \bibfield  {author} {\bibinfo {author} {\bibfnamefont {V.~J.}\ \bibnamefont
  {Emery}}, \bibinfo {author} {\bibfnamefont {S.~A.}\ \bibnamefont {Kivelson}},
  \ and\ \bibinfo {author} {\bibfnamefont {J.~M.}\ \bibnamefont {Tranquada}},\
  }\href {http://www.pnas.org/content/96/16/8814.full} {\bibfield  {journal}
  {\bibinfo  {journal} {Proc. Natl. Acad. Sci.}\ }\textbf {\bibinfo {volume}
  {96}},\ \bibinfo {pages} {8814} (\bibinfo {year} {1999})}\BibitemShut
  {NoStop}%
\bibitem [{\citenamefont {He}\ \emph {et~al.}(2011)\citenamefont {He},
  \citenamefont {Hashimoto}, \citenamefont {Karapetyan}, \citenamefont
  {Koralek}, \citenamefont {Hinton}, \citenamefont {Testaud}, \citenamefont
  {Nathan}, \citenamefont {Yoshida}, \citenamefont {Yao}, \citenamefont
  {Tanaka}, \citenamefont {Meevasana}, \citenamefont {Moore}, \citenamefont
  {Lu}, \citenamefont {Mo}, \citenamefont {Ishikado}, \citenamefont {Eisaki},
  \citenamefont {Hussain}, \citenamefont {Devereaux}, \citenamefont {Kivelson},
  \citenamefont {Orenstein}, \citenamefont {Kapitulnik},\ and\ \citenamefont
  {Shen}}]{He2011c}%
  \BibitemOpen
  \bibfield  {author} {\bibinfo {author} {\bibfnamefont {R.-H.}\ \bibnamefont
  {He}}, \bibinfo {author} {\bibfnamefont {M.}~\bibnamefont {Hashimoto}},
  \bibinfo {author} {\bibfnamefont {H.}~\bibnamefont {Karapetyan}}, \bibinfo
  {author} {\bibfnamefont {J.~D.}\ \bibnamefont {Koralek}}, \bibinfo {author}
  {\bibfnamefont {J.~P.}\ \bibnamefont {Hinton}}, \bibinfo {author}
  {\bibfnamefont {J.~P.}\ \bibnamefont {Testaud}}, \bibinfo {author}
  {\bibfnamefont {V.}~\bibnamefont {Nathan}}, \bibinfo {author} {\bibfnamefont
  {Y.}~\bibnamefont {Yoshida}}, \bibinfo {author} {\bibfnamefont
  {H.}~\bibnamefont {Yao}}, \bibinfo {author} {\bibfnamefont {K.}~\bibnamefont
  {Tanaka}}, \bibinfo {author} {\bibfnamefont {W.}~\bibnamefont {Meevasana}},
  \bibinfo {author} {\bibfnamefont {R.~G.}\ \bibnamefont {Moore}}, \bibinfo
  {author} {\bibfnamefont {D.~H.}\ \bibnamefont {Lu}}, \bibinfo {author}
  {\bibfnamefont {S.-K.}\ \bibnamefont {Mo}}, \bibinfo {author} {\bibfnamefont
  {M.}~\bibnamefont {Ishikado}}, \bibinfo {author} {\bibfnamefont
  {H.}~\bibnamefont {Eisaki}}, \bibinfo {author} {\bibfnamefont
  {Z.}~\bibnamefont {Hussain}}, \bibinfo {author} {\bibfnamefont {T.~P.}\
  \bibnamefont {Devereaux}}, \bibinfo {author} {\bibfnamefont {S.~A.}\
  \bibnamefont {Kivelson}}, \bibinfo {author} {\bibfnamefont {J.}~\bibnamefont
  {Orenstein}}, \bibinfo {author} {\bibfnamefont {A.}~\bibnamefont
  {Kapitulnik}}, \ and\ \bibinfo {author} {\bibfnamefont {Z.-X.}\ \bibnamefont
  {Shen}},\ }\href {\doibase 10.1126/science.1198415} {\bibfield  {journal}
  {\bibinfo  {journal} {Science}\ }\textbf {\bibinfo {volume} {331}},\ \bibinfo
  {pages} {1579} (\bibinfo {year} {2011})}\BibitemShut {NoStop}%
\bibitem [{\citenamefont {Xia}\ \emph {et~al.}(2008)\citenamefont {Xia},
  \citenamefont {Schemm}, \citenamefont {Deutscher}, \citenamefont {Kivelson},
  \citenamefont {Bonn}, \citenamefont {Hardy}, \citenamefont {Liang},
  \citenamefont {Siemons}, \citenamefont {Koster}, \citenamefont {Fejer},\ and\
  \citenamefont {Kapitulnik}}]{Xia2008}%
  \BibitemOpen
  \bibfield  {author} {\bibinfo {author} {\bibfnamefont {J.}~\bibnamefont
  {Xia}}, \bibinfo {author} {\bibfnamefont {E.}~\bibnamefont {Schemm}},
  \bibinfo {author} {\bibfnamefont {G.}~\bibnamefont {Deutscher}}, \bibinfo
  {author} {\bibfnamefont {S.~A.}\ \bibnamefont {Kivelson}}, \bibinfo {author}
  {\bibfnamefont {D.~A.}\ \bibnamefont {Bonn}}, \bibinfo {author}
  {\bibfnamefont {W.~N.}\ \bibnamefont {Hardy}}, \bibinfo {author}
  {\bibfnamefont {R.}~\bibnamefont {Liang}}, \bibinfo {author} {\bibfnamefont
  {W.}~\bibnamefont {Siemons}}, \bibinfo {author} {\bibfnamefont
  {G.}~\bibnamefont {Koster}}, \bibinfo {author} {\bibfnamefont {M.~M.}\
  \bibnamefont {Fejer}}, \ and\ \bibinfo {author} {\bibfnamefont
  {A.}~\bibnamefont {Kapitulnik}},\ }\href {\doibase
  10.1103/PhysRevLett.100.127002} {\bibfield  {journal} {\bibinfo  {journal}
  {Phys. Rev. Lett.}\ }\textbf {\bibinfo {volume} {100}},\ \bibinfo {pages} {3}
  (\bibinfo {year} {2008})}\BibitemShut {NoStop}%
\bibitem [{\citenamefont {Fujita}\ \emph {et~al.}(2004)\citenamefont {Fujita},
  \citenamefont {Goka}, \citenamefont {Yamada}, \citenamefont {Tranquada},\
  and\ \citenamefont {Regnault}}]{Fujita2004}%
  \BibitemOpen
  \bibfield  {author} {\bibinfo {author} {\bibfnamefont {M.}~\bibnamefont
  {Fujita}}, \bibinfo {author} {\bibfnamefont {H.}~\bibnamefont {Goka}},
  \bibinfo {author} {\bibfnamefont {K.}~\bibnamefont {Yamada}}, \bibinfo
  {author} {\bibfnamefont {J.~M.}\ \bibnamefont {Tranquada}}, \ and\ \bibinfo
  {author} {\bibfnamefont {L.~P.}\ \bibnamefont {Regnault}},\ }\href {\doibase
  10.1103/PhysRevB.70.104517} {\bibfield  {journal} {\bibinfo  {journal} {Phys.
  Rev. B}\ }\textbf {\bibinfo {volume} {70}},\ \bibinfo {pages} {104517}
  (\bibinfo {year} {2004})}\BibitemShut {NoStop}%
\bibitem [{\citenamefont {H\"{u}cker}\ \emph {et~al.}(2011)\citenamefont
  {H\"{u}cker}, \citenamefont {v.~Zimmermann}, \citenamefont {Gu},
  \citenamefont {Xu}, \citenamefont {Wen}, \citenamefont {Xu}, \citenamefont
  {Kang}, \citenamefont {Zheludev},\ and\ \citenamefont
  {Tranquada}}]{Hucker2011a}%
  \BibitemOpen
  \bibfield  {author} {\bibinfo {author} {\bibfnamefont {M.}~\bibnamefont
  {H\"{u}cker}}, \bibinfo {author} {\bibfnamefont {M.}~\bibnamefont
  {v.~Zimmermann}}, \bibinfo {author} {\bibfnamefont {G.~D.}\ \bibnamefont
  {Gu}}, \bibinfo {author} {\bibfnamefont {Z.~J.}\ \bibnamefont {Xu}}, \bibinfo
  {author} {\bibfnamefont {J.~S.}\ \bibnamefont {Wen}}, \bibinfo {author}
  {\bibfnamefont {G.}~\bibnamefont {Xu}}, \bibinfo {author} {\bibfnamefont
  {H.~J.}\ \bibnamefont {Kang}}, \bibinfo {author} {\bibfnamefont
  {A.}~\bibnamefont {Zheludev}}, \ and\ \bibinfo {author} {\bibfnamefont
  {J.~M.}\ \bibnamefont {Tranquada}},\ }\href {\doibase
  10.1103/PhysRevB.83.104506} {\bibfield  {journal} {\bibinfo  {journal} {Phys.
  Rev. B}\ }\textbf {\bibinfo {volume} {83}},\ \bibinfo {pages} {104506}
  (\bibinfo {year} {2011})}\BibitemShut {NoStop}%
\bibitem [{\citenamefont {Li}\ \emph {et~al.}(2011)\citenamefont {Li},
  \citenamefont {Alidoust}, \citenamefont {Tranquada}, \citenamefont {Gu},\
  and\ \citenamefont {Ong}}]{Li2011d}%
  \BibitemOpen
  \bibfield  {author} {\bibinfo {author} {\bibfnamefont {L.}~\bibnamefont
  {Li}}, \bibinfo {author} {\bibfnamefont {N.}~\bibnamefont {Alidoust}},
  \bibinfo {author} {\bibfnamefont {J.}~\bibnamefont {Tranquada}}, \bibinfo
  {author} {\bibfnamefont {G.}~\bibnamefont {Gu}}, \ and\ \bibinfo {author}
  {\bibfnamefont {N.~P.}\ \bibnamefont {Ong}},\ }\href {\doibase
  10.1103/PhysRevLett.107.277001} {\bibfield  {journal} {\bibinfo  {journal}
  {Phys. Rev. Lett.}\ }\textbf {\bibinfo {volume} {107}},\ \bibinfo {pages} {1}
  (\bibinfo {year} {2011})}\BibitemShut {NoStop}%
\bibitem [{\citenamefont {Karapetyan}\ \emph {et~al.}(2012)\citenamefont
  {Karapetyan}, \citenamefont {H\"{u}cker}, \citenamefont {Gu}, \citenamefont
  {Tranquada}, \citenamefont {Fejer}, \citenamefont {Xia},\ and\ \citenamefont
  {Kapitulnik}}]{Karapetyan2012}%
  \BibitemOpen
  \bibfield  {author} {\bibinfo {author} {\bibfnamefont {H.}~\bibnamefont
  {Karapetyan}}, \bibinfo {author} {\bibfnamefont {M.}~\bibnamefont
  {H\"{u}cker}}, \bibinfo {author} {\bibfnamefont {G.}~\bibnamefont {Gu}},
  \bibinfo {author} {\bibfnamefont {J.}~\bibnamefont {Tranquada}}, \bibinfo
  {author} {\bibfnamefont {M.}~\bibnamefont {Fejer}}, \bibinfo {author}
  {\bibfnamefont {J.}~\bibnamefont {Xia}}, \ and\ \bibinfo {author}
  {\bibfnamefont {A.}~\bibnamefont {Kapitulnik}},\ }\href {\doibase
  10.1103/PhysRevLett.109.147001} {\bibfield  {journal} {\bibinfo  {journal}
  {Phys. Rev. Lett.}\ }\textbf {\bibinfo {volume} {109}},\ \bibinfo {pages} {1}
  (\bibinfo {year} {2012})}\BibitemShut {NoStop}%
\bibitem [{\citenamefont {Karapetyan}\ \emph {et~al.}(2014)\citenamefont
  {Karapetyan}, \citenamefont {Xia}, \citenamefont {H\"{u}cker}, \citenamefont
  {Gu}, \citenamefont {Tranquada}, \citenamefont {Fejer},\ and\ \citenamefont
  {Kapitulnik}}]{Karapetyan2014}%
  \BibitemOpen
  \bibfield  {author} {\bibinfo {author} {\bibfnamefont {H.}~\bibnamefont
  {Karapetyan}}, \bibinfo {author} {\bibfnamefont {J.}~\bibnamefont {Xia}},
  \bibinfo {author} {\bibfnamefont {M.}~\bibnamefont {H\"{u}cker}}, \bibinfo
  {author} {\bibfnamefont {G.~D.}\ \bibnamefont {Gu}}, \bibinfo {author}
  {\bibfnamefont {J.~M.}\ \bibnamefont {Tranquada}}, \bibinfo {author}
  {\bibfnamefont {M.~M.}\ \bibnamefont {Fejer}}, \ and\ \bibinfo {author}
  {\bibfnamefont {A.}~\bibnamefont {Kapitulnik}},\ }\href {\doibase
  10.1103/PhysRevLett.112.047003} {\bibfield  {journal} {\bibinfo  {journal}
  {Phys. Rev. Lett.}\ }\textbf {\bibinfo {volume} {112}},\ \bibinfo {pages}
  {047003} (\bibinfo {year} {2014})}\BibitemShut {NoStop}%
\bibitem [{\citenamefont {Hosur}\ \emph {et~al.}(2013)\citenamefont {Hosur},
  \citenamefont {Kapitulnik}, \citenamefont {Kivelson}, \citenamefont
  {Orenstein},\ and\ \citenamefont {Raghu}}]{Hosur2013}%
  \BibitemOpen
  \bibfield  {author} {\bibinfo {author} {\bibfnamefont {P.}~\bibnamefont
  {Hosur}}, \bibinfo {author} {\bibfnamefont {A.}~\bibnamefont {Kapitulnik}},
  \bibinfo {author} {\bibfnamefont {S.~A.}\ \bibnamefont {Kivelson}}, \bibinfo
  {author} {\bibfnamefont {J.}~\bibnamefont {Orenstein}}, \ and\ \bibinfo
  {author} {\bibfnamefont {S.}~\bibnamefont {Raghu}},\ }\href {\doibase
  10.1103/PhysRevB.87.115116} {\bibfield  {journal} {\bibinfo  {journal} {Phys.
  Rev. B}\ }\textbf {\bibinfo {volume} {87}},\ \bibinfo {pages} {115116}
  (\bibinfo {year} {2013})}\BibitemShut {NoStop}%
\bibitem [{\citenamefont {Valla}\ \emph {et~al.}(2006)\citenamefont {Valla},
  \citenamefont {Fedorov}, \citenamefont {Lee}, \citenamefont {Davis},\ and\
  \citenamefont {Gu}}]{Valla2006}%
  \BibitemOpen
  \bibfield  {author} {\bibinfo {author} {\bibfnamefont {T.}~\bibnamefont
  {Valla}}, \bibinfo {author} {\bibfnamefont {a.~V.}\ \bibnamefont {Fedorov}},
  \bibinfo {author} {\bibfnamefont {J.}~\bibnamefont {Lee}}, \bibinfo {author}
  {\bibfnamefont {J.~C.}\ \bibnamefont {Davis}}, \ and\ \bibinfo {author}
  {\bibfnamefont {G.~D.}\ \bibnamefont {Gu}},\ }\href {\doibase
  10.1126/science.1134742} {\bibfield  {journal} {\bibinfo  {journal}
  {Science}\ }\textbf {\bibinfo {volume} {314}},\ \bibinfo {pages} {1914}
  (\bibinfo {year} {2006})}\BibitemShut {NoStop}%
\bibitem [{\citenamefont {He}\ \emph {et~al.}(2008)\citenamefont {He},
  \citenamefont {Tanaka}, \citenamefont {Mo}, \citenamefont {Sasagawa},
  \citenamefont {Fujita}, \citenamefont {Adachi}, \citenamefont {Mannella},
  \citenamefont {Yamada}, \citenamefont {Koike}, \citenamefont {Hussain},\ and\
  \citenamefont {Shen}}]{He2008}%
  \BibitemOpen
  \bibfield  {author} {\bibinfo {author} {\bibfnamefont {R.-H.}\ \bibnamefont
  {He}}, \bibinfo {author} {\bibfnamefont {K.}~\bibnamefont {Tanaka}}, \bibinfo
  {author} {\bibfnamefont {S.-K.}\ \bibnamefont {Mo}}, \bibinfo {author}
  {\bibfnamefont {T.}~\bibnamefont {Sasagawa}}, \bibinfo {author}
  {\bibfnamefont {M.}~\bibnamefont {Fujita}}, \bibinfo {author} {\bibfnamefont
  {T.}~\bibnamefont {Adachi}}, \bibinfo {author} {\bibfnamefont
  {N.}~\bibnamefont {Mannella}}, \bibinfo {author} {\bibfnamefont
  {K.}~\bibnamefont {Yamada}}, \bibinfo {author} {\bibfnamefont
  {Y.}~\bibnamefont {Koike}}, \bibinfo {author} {\bibfnamefont
  {Z.}~\bibnamefont {Hussain}}, \ and\ \bibinfo {author} {\bibfnamefont
  {Z.-X.}\ \bibnamefont {Shen}},\ }\href {\doibase 10.1038/nphys1159}
  {\bibfield  {journal} {\bibinfo  {journal} {Nat. Phys.}\ }\textbf {\bibinfo
  {volume} {5}},\ \bibinfo {pages} {119} (\bibinfo {year} {2008})}\BibitemShut
  {NoStop}%
\bibitem [{\citenamefont {Li}\ \emph {et~al.}(2007{\natexlab{a}})\citenamefont
  {Li}, \citenamefont {H\"{u}cker}, \citenamefont {Gu}, \citenamefont
  {Tsvelik},\ and\ \citenamefont {Tranquada}}]{Li2007c}%
  \BibitemOpen
  \bibfield  {author} {\bibinfo {author} {\bibfnamefont {Q.}~\bibnamefont
  {Li}}, \bibinfo {author} {\bibfnamefont {M.}~\bibnamefont {H\"{u}cker}},
  \bibinfo {author} {\bibfnamefont {G.}~\bibnamefont {Gu}}, \bibinfo {author}
  {\bibfnamefont {A.}~\bibnamefont {Tsvelik}}, \ and\ \bibinfo {author}
  {\bibfnamefont {J.}~\bibnamefont {Tranquada}},\ }\href {\doibase
  10.1103/PhysRevLett.99.067001} {\bibfield  {journal} {\bibinfo  {journal}
  {Phys. Rev. Lett.}\ }\textbf {\bibinfo {volume} {99}},\ \bibinfo {pages}
  {067001} (\bibinfo {year} {2007}{\natexlab{a}})}\BibitemShut {NoStop}%
\bibitem [{\citenamefont {Tranquada}\ \emph {et~al.}(2008)\citenamefont
  {Tranquada}, \citenamefont {Gu}, \citenamefont {H\"{u}cker}, \citenamefont
  {Jie}, \citenamefont {Kang}, \citenamefont {Klingeler}, \citenamefont {Li},
  \citenamefont {Tristan}, \citenamefont {Wen}, \citenamefont {Xu},
  \citenamefont {Xu}, \citenamefont {Zhou},\ and\ \citenamefont {von
  Zimmermann}}]{Tranquada2008}%
  \BibitemOpen
  \bibfield  {author} {\bibinfo {author} {\bibfnamefont {J.~M.}\ \bibnamefont
  {Tranquada}}, \bibinfo {author} {\bibfnamefont {G.~D.}\ \bibnamefont {Gu}},
  \bibinfo {author} {\bibfnamefont {M.}~\bibnamefont {H\"{u}cker}}, \bibinfo
  {author} {\bibfnamefont {Q.}~\bibnamefont {Jie}}, \bibinfo {author}
  {\bibfnamefont {H.-J.}\ \bibnamefont {Kang}}, \bibinfo {author}
  {\bibfnamefont {R.}~\bibnamefont {Klingeler}}, \bibinfo {author}
  {\bibfnamefont {Q.}~\bibnamefont {Li}}, \bibinfo {author} {\bibfnamefont
  {N.}~\bibnamefont {Tristan}}, \bibinfo {author} {\bibfnamefont {J.~S.}\
  \bibnamefont {Wen}}, \bibinfo {author} {\bibfnamefont {G.}~\bibnamefont
  {Xu}}, \bibinfo {author} {\bibfnamefont {Z.~A.}\ \bibnamefont {Xu}}, \bibinfo
  {author} {\bibfnamefont {J.}~\bibnamefont {Zhou}}, \ and\ \bibinfo {author}
  {\bibfnamefont {M.}~\bibnamefont {von Zimmermann}},\ }\href {\doibase
  10.1103/PhysRevB.78.174529} {\bibfield  {journal} {\bibinfo  {journal} {Phys.
  Rev. B}\ }\textbf {\bibinfo {volume} {78}},\ \bibinfo {pages} {174529}
  (\bibinfo {year} {2008})}\BibitemShut {NoStop}%
\bibitem [{\citenamefont {Orenstein}\ and\ \citenamefont
  {Moore}(2013)}]{Orenstein2013}%
  \BibitemOpen
  \bibfield  {author} {\bibinfo {author} {\bibfnamefont {J.}~\bibnamefont
  {Orenstein}}\ and\ \bibinfo {author} {\bibfnamefont {J.~E.}\ \bibnamefont
  {Moore}},\ }\href {\doibase 10.1103/PhysRevB.87.165110} {\bibfield  {journal}
  {\bibinfo  {journal} {Phys. Rev. B}\ }\textbf {\bibinfo {volume} {87}},\
  \bibinfo {pages} {165110} (\bibinfo {year} {2013})}\BibitemShut {NoStop}%
\bibitem [{\citenamefont {Chakravarty}(2014)}]{Chakravarty2014}%
  \BibitemOpen
  \bibfield  {author} {\bibinfo {author} {\bibfnamefont {S.}~\bibnamefont
  {Chakravarty}},\ }\href {\doibase 10.1103/PhysRevB.89.087101} {\bibfield
  {journal} {\bibinfo  {journal} {Phys. Rev. B}\ }\textbf {\bibinfo {volume}
  {89}},\ \bibinfo {pages} {087101} (\bibinfo {year} {2014})}\BibitemShut
  {NoStop}%
\bibitem [{\citenamefont {Armitage}(2014)}]{Armitage2014}%
  \BibitemOpen
  \bibfield  {author} {\bibinfo {author} {\bibfnamefont {N.~P.}\ \bibnamefont
  {Armitage}},\ }\href {\doibase 10.1103/PhysRevB.90.035135} {\bibfield
  {journal} {\bibinfo  {journal} {Phys. Rev. B}\ }\textbf {\bibinfo {volume}
  {90}},\ \bibinfo {pages} {035135} (\bibinfo {year} {2014})}\BibitemShut
  {NoStop}%
\bibitem [{\citenamefont {Hosur}\ \emph {et~al.}(2015)\citenamefont {Hosur},
  \citenamefont {Kapitulnik}, \citenamefont {Kivelson}, \citenamefont
  {Orenstein}, \citenamefont {Raghu}, \citenamefont {Cho},\ and\ \citenamefont
  {Fried}}]{Hosur2015}%
  \BibitemOpen
  \bibfield  {author} {\bibinfo {author} {\bibfnamefont {P.}~\bibnamefont
  {Hosur}}, \bibinfo {author} {\bibfnamefont {A.}~\bibnamefont {Kapitulnik}},
  \bibinfo {author} {\bibfnamefont {S.~A.}\ \bibnamefont {Kivelson}}, \bibinfo
  {author} {\bibfnamefont {J.}~\bibnamefont {Orenstein}}, \bibinfo {author}
  {\bibfnamefont {S.}~\bibnamefont {Raghu}}, \bibinfo {author} {\bibfnamefont
  {W.}~\bibnamefont {Cho}}, \ and\ \bibinfo {author} {\bibfnamefont
  {A.}~\bibnamefont {Fried}},\ }\href {\doibase 10.1103/PhysRevB.91.039908}
  {\bibfield  {journal} {\bibinfo  {journal} {Physical Review B}\ }\textbf
  {\bibinfo {volume} {91}},\ \bibinfo {pages} {039908} (\bibinfo {year}
  {2015})}\BibitemShut {NoStop}%
\bibitem [{SOM()}]{SOM}%
  \BibitemOpen
  \href@noop {} {\emph {\bibinfo {title} {Supplementary
  Materials}}}\BibitemShut {NoStop}%
\bibitem [{\citenamefont {Ito}\ \emph {et~al.}(2013)\citenamefont {Ito},
  \citenamefont {Ushiyama}, \citenamefont {Yanagisawa}, \citenamefont
  {Tomioka}, \citenamefont {Shindo},\ and\ \citenamefont {Yanase}}]{Ito2013}%
  \BibitemOpen
  \bibfield  {author} {\bibinfo {author} {\bibfnamefont {T.}~\bibnamefont
  {Ito}}, \bibinfo {author} {\bibfnamefont {T.}~\bibnamefont {Ushiyama}},
  \bibinfo {author} {\bibfnamefont {Y.}~\bibnamefont {Yanagisawa}}, \bibinfo
  {author} {\bibfnamefont {Y.}~\bibnamefont {Tomioka}}, \bibinfo {author}
  {\bibfnamefont {I.}~\bibnamefont {Shindo}}, \ and\ \bibinfo {author}
  {\bibfnamefont {A.}~\bibnamefont {Yanase}},\ }\href {\doibase
  10.1016/j.jcrysgro.2012.10.059} {\bibfield  {journal} {\bibinfo  {journal}
  {J. Cryst. Growth}\ }\textbf {\bibinfo {volume} {363}},\ \bibinfo {pages}
  {264} (\bibinfo {year} {2013})}\BibitemShut {NoStop}%
\bibitem [{\citenamefont {H\"{u}cker}\ \emph {et~al.}(2008)\citenamefont
  {H\"{u}cker}, \citenamefont {Gu},\ and\ \citenamefont
  {Tranquada}}]{Hucker2008}%
  \BibitemOpen
  \bibfield  {author} {\bibinfo {author} {\bibfnamefont {M.}~\bibnamefont
  {H\"{u}cker}}, \bibinfo {author} {\bibfnamefont {G.}~\bibnamefont {Gu}}, \
  and\ \bibinfo {author} {\bibfnamefont {J.}~\bibnamefont {Tranquada}},\ }\href
  {\doibase 10.1103/PhysRevB.78.214507} {\bibfield  {journal} {\bibinfo
  {journal} {Phys. Rev. B}\ }\textbf {\bibinfo {volume} {78}},\ \bibinfo
  {pages} {214507} (\bibinfo {year} {2008})}\BibitemShut {NoStop}%
\bibitem [{\citenamefont {H\"{u}cker}\ \emph {et~al.}(2010)\citenamefont
  {H\"{u}cker}, \citenamefont {v.~Zimmermann}, \citenamefont {Debessai},
  \citenamefont {Schilling}, \citenamefont {Tranquada},\ and\ \citenamefont
  {Gu}}]{Hucker2010}%
  \BibitemOpen
  \bibfield  {author} {\bibinfo {author} {\bibfnamefont {M.}~\bibnamefont
  {H\"{u}cker}}, \bibinfo {author} {\bibfnamefont {M.}~\bibnamefont
  {v.~Zimmermann}}, \bibinfo {author} {\bibfnamefont {M.}~\bibnamefont
  {Debessai}}, \bibinfo {author} {\bibfnamefont {J.~S.}\ \bibnamefont
  {Schilling}}, \bibinfo {author} {\bibfnamefont {J.~M.}\ \bibnamefont
  {Tranquada}}, \ and\ \bibinfo {author} {\bibfnamefont {G.~D.}\ \bibnamefont
  {Gu}},\ }\href {\doibase 10.1103/PhysRevLett.104.057004} {\bibfield
  {journal} {\bibinfo  {journal} {Phys. Rev. Lett.}\ }\textbf {\bibinfo
  {volume} {104}},\ \bibinfo {pages} {057004} (\bibinfo {year}
  {2010})}\BibitemShut {NoStop}%
\bibitem [{\citenamefont {Nakamura}\ and\ \citenamefont
  {Uchida}(1993)}]{Nakamura1993}%
  \BibitemOpen
  \bibfield  {author} {\bibinfo {author} {\bibfnamefont {Y.}~\bibnamefont
  {Nakamura}}\ and\ \bibinfo {author} {\bibfnamefont {S.}~\bibnamefont
  {Uchida}},\ }\href {\doibase 10.1103/PhysRevB.47.8369} {\bibfield  {journal}
  {\bibinfo  {journal} {Phys. Rev. B}\ }\textbf {\bibinfo {volume} {47}},\
  \bibinfo {pages} {8369} (\bibinfo {year} {1993})}\BibitemShut {NoStop}%
\bibitem [{\citenamefont {Wang}\ \emph {et~al.}(2006)\citenamefont {Wang},
  \citenamefont {Li},\ and\ \citenamefont {Ong}}]{Wang2006}%
  \BibitemOpen
  \bibfield  {author} {\bibinfo {author} {\bibfnamefont {Y.}~\bibnamefont
  {Wang}}, \bibinfo {author} {\bibfnamefont {L.}~\bibnamefont {Li}}, \ and\
  \bibinfo {author} {\bibfnamefont {N.~P.}\ \bibnamefont {Ong}},\ }\href
  {\doibase 10.1103/PhysRevB.73.024510} {\bibfield  {journal} {\bibinfo
  {journal} {Phys. Rev. B}\ }\textbf {\bibinfo {volume} {73}},\ \bibinfo
  {pages} {024510} (\bibinfo {year} {2006})}\BibitemShut {NoStop}%
\bibitem [{\citenamefont {Li}\ \emph {et~al.}(2007{\natexlab{b}})\citenamefont
  {Li}, \citenamefont {Checkelsky}, \citenamefont {Komiya}, \citenamefont
  {Ando},\ and\ \citenamefont {Ong}}]{Li2007a}%
  \BibitemOpen
  \bibfield  {author} {\bibinfo {author} {\bibfnamefont {L.}~\bibnamefont
  {Li}}, \bibinfo {author} {\bibfnamefont {J.~G.}\ \bibnamefont {Checkelsky}},
  \bibinfo {author} {\bibfnamefont {S.}~\bibnamefont {Komiya}}, \bibinfo
  {author} {\bibfnamefont {Y.}~\bibnamefont {Ando}}, \ and\ \bibinfo {author}
  {\bibfnamefont {N.~P.}\ \bibnamefont {Ong}},\ }\href {\doibase
  10.1038/nphys563} {\bibfield  {journal} {\bibinfo  {journal} {Nat. Phys.}\
  }\textbf {\bibinfo {volume} {3}},\ \bibinfo {pages} {311} (\bibinfo {year}
  {2007}{\natexlab{b}})}\BibitemShut {NoStop}%
\bibitem [{\citenamefont {Ussishkin}\ \emph {et~al.}(2002)\citenamefont
  {Ussishkin}, \citenamefont {Sondhi},\ and\ \citenamefont
  {Huse}}]{Ussishkin2002}%
  \BibitemOpen
  \bibfield  {author} {\bibinfo {author} {\bibfnamefont {I.}~\bibnamefont
  {Ussishkin}}, \bibinfo {author} {\bibfnamefont {S.}~\bibnamefont {Sondhi}}, \
  and\ \bibinfo {author} {\bibfnamefont {D.}~\bibnamefont {Huse}},\ }\href
  {\doibase 10.1103/PhysRevLett.89.287001} {\bibfield  {journal} {\bibinfo
  {journal} {Phys. Rev. Lett.}\ }\textbf {\bibinfo {volume} {89}},\ \bibinfo
  {pages} {287001} (\bibinfo {year} {2002})}\BibitemShut {NoStop}%
\bibitem [{\citenamefont {Cyr-Choini\`{e}re}\ \emph {et~al.}(2009)\citenamefont
  {Cyr-Choini\`{e}re}, \citenamefont {Daou}, \citenamefont {Lalibert\'{e}},
  \citenamefont {LeBoeuf}, \citenamefont {Doiron-Leyraud}, \citenamefont
  {Chang}, \citenamefont {Yan}, \citenamefont {Cheng}, \citenamefont {Zhou},
  \citenamefont {Goodenough}, \citenamefont {Pyon}, \citenamefont {Takayama},
  \citenamefont {Takagi}, \citenamefont {Tanaka},\ and\ \citenamefont
  {Taillefer}}]{Cyr-Choiniere2009}%
  \BibitemOpen
  \bibfield  {author} {\bibinfo {author} {\bibfnamefont {O.}~\bibnamefont
  {Cyr-Choini\`{e}re}}, \bibinfo {author} {\bibfnamefont {R.}~\bibnamefont
  {Daou}}, \bibinfo {author} {\bibfnamefont {F.}~\bibnamefont {Lalibert\'{e}}},
  \bibinfo {author} {\bibfnamefont {D.}~\bibnamefont {LeBoeuf}}, \bibinfo
  {author} {\bibfnamefont {N.}~\bibnamefont {Doiron-Leyraud}}, \bibinfo
  {author} {\bibfnamefont {J.}~\bibnamefont {Chang}}, \bibinfo {author}
  {\bibfnamefont {J.-Q.}\ \bibnamefont {Yan}}, \bibinfo {author} {\bibfnamefont
  {J.-G.}\ \bibnamefont {Cheng}}, \bibinfo {author} {\bibfnamefont {J.-S.}\
  \bibnamefont {Zhou}}, \bibinfo {author} {\bibfnamefont {J.~B.}\ \bibnamefont
  {Goodenough}}, \bibinfo {author} {\bibfnamefont {S.}~\bibnamefont {Pyon}},
  \bibinfo {author} {\bibfnamefont {T.}~\bibnamefont {Takayama}}, \bibinfo
  {author} {\bibfnamefont {H.}~\bibnamefont {Takagi}}, \bibinfo {author}
  {\bibfnamefont {Y.}~\bibnamefont {Tanaka}}, \ and\ \bibinfo {author}
  {\bibfnamefont {L.}~\bibnamefont {Taillefer}},\ }\href {\doibase
  10.1038/nature07931} {\bibfield  {journal} {\bibinfo  {journal} {Nature}\
  }\textbf {\bibinfo {volume} {458}},\ \bibinfo {pages} {743} (\bibinfo {year}
  {2009})}\BibitemShut {NoStop}%
\bibitem [{\citenamefont {Chang}\ \emph {et~al.}(2010)\citenamefont {Chang},
  \citenamefont {Daou}, \citenamefont {Proust}, \citenamefont {LeBoeuf},
  \citenamefont {Doiron-Leyraud}, \citenamefont {Lalibert\'{e}}, \citenamefont
  {Pingault}, \citenamefont {Ramshaw}, \citenamefont {Liang}, \citenamefont
  {Bonn}, \citenamefont {Hardy}, \citenamefont {Takagi}, \citenamefont
  {Antunes}, \citenamefont {Sheikin}, \citenamefont {Behnia},\ and\
  \citenamefont {Taillefer}}]{Chang2010}%
  \BibitemOpen
  \bibfield  {author} {\bibinfo {author} {\bibfnamefont {J.}~\bibnamefont
  {Chang}}, \bibinfo {author} {\bibfnamefont {R.}~\bibnamefont {Daou}},
  \bibinfo {author} {\bibfnamefont {C.}~\bibnamefont {Proust}}, \bibinfo
  {author} {\bibfnamefont {D.}~\bibnamefont {LeBoeuf}}, \bibinfo {author}
  {\bibfnamefont {N.}~\bibnamefont {Doiron-Leyraud}}, \bibinfo {author}
  {\bibfnamefont {F.}~\bibnamefont {Lalibert\'{e}}}, \bibinfo {author}
  {\bibfnamefont {B.}~\bibnamefont {Pingault}}, \bibinfo {author}
  {\bibfnamefont {B.~J.}\ \bibnamefont {Ramshaw}}, \bibinfo {author}
  {\bibfnamefont {R.}~\bibnamefont {Liang}}, \bibinfo {author} {\bibfnamefont
  {D.~A.}\ \bibnamefont {Bonn}}, \bibinfo {author} {\bibfnamefont {W.~N.}\
  \bibnamefont {Hardy}}, \bibinfo {author} {\bibfnamefont {H.}~\bibnamefont
  {Takagi}}, \bibinfo {author} {\bibfnamefont {A.~B.}\ \bibnamefont {Antunes}},
  \bibinfo {author} {\bibfnamefont {I.}~\bibnamefont {Sheikin}}, \bibinfo
  {author} {\bibfnamefont {K.}~\bibnamefont {Behnia}}, \ and\ \bibinfo {author}
  {\bibfnamefont {L.}~\bibnamefont {Taillefer}},\ }\href {\doibase
  10.1103/PhysRevLett.104.057005} {\bibfield  {journal} {\bibinfo  {journal}
  {Phys. Rev. Lett.}\ }\textbf {\bibinfo {volume} {104}},\ \bibinfo {pages} {3}
  (\bibinfo {year} {2010})}\BibitemShut {NoStop}%
\bibitem [{\citenamefont {Serbyn}\ \emph {et~al.}(2009)\citenamefont {Serbyn},
  \citenamefont {Skvortsov}, \citenamefont {Varlamov},\ and\ \citenamefont
  {Galitski}}]{Serbyn2009}%
  \BibitemOpen
  \bibfield  {author} {\bibinfo {author} {\bibfnamefont {M.}~\bibnamefont
  {Serbyn}}, \bibinfo {author} {\bibfnamefont {M.}~\bibnamefont {Skvortsov}},
  \bibinfo {author} {\bibfnamefont {A.}~\bibnamefont {Varlamov}}, \ and\
  \bibinfo {author} {\bibfnamefont {V.}~\bibnamefont {Galitski}},\ }\href
  {\doibase 10.1103/PhysRevLett.102.067001} {\bibfield  {journal} {\bibinfo
  {journal} {Phys. Rev. Lett.}\ }\textbf {\bibinfo {volume} {102}},\ \bibinfo
  {pages} {067001} (\bibinfo {year} {2009})}\BibitemShut {NoStop}%
\bibitem [{\citenamefont {Michaeli}\ and\ \citenamefont
  {Finkel'stein}(2009)}]{Michaeli2009}%
  \BibitemOpen
  \bibfield  {author} {\bibinfo {author} {\bibfnamefont {K.}~\bibnamefont
  {Michaeli}}\ and\ \bibinfo {author} {\bibfnamefont {A.~M.}\ \bibnamefont
  {Finkel'stein}},\ }\href {\doibase 10.1209/0295-5075/86/27007} {\bibfield
  {journal} {\bibinfo  {journal} {Europhys. Lett.}\ }\textbf {\bibinfo {volume}
  {86}},\ \bibinfo {pages} {27007} (\bibinfo {year} {2009})}\BibitemShut
  {NoStop}%
\bibitem [{\citenamefont {Xu}\ \emph {et~al.}(2000)\citenamefont {Xu},
  \citenamefont {Ong}, \citenamefont {Wang}, \citenamefont {Kakeshita},\ and\
  \citenamefont {Uchida}}]{Xu2000}%
  \BibitemOpen
  \bibfield  {author} {\bibinfo {author} {\bibfnamefont {Z.~A.}\ \bibnamefont
  {Xu}}, \bibinfo {author} {\bibfnamefont {N.~P.}\ \bibnamefont {Ong}},
  \bibinfo {author} {\bibfnamefont {Y.}~\bibnamefont {Wang}}, \bibinfo {author}
  {\bibfnamefont {T.}~\bibnamefont {Kakeshita}}, \ and\ \bibinfo {author}
  {\bibfnamefont {S.}~\bibnamefont {Uchida}},\ }\href {\doibase
  10.1038/35020016} {\bibfield  {journal} {\bibinfo  {journal} {Nature}\
  }\textbf {\bibinfo {volume} {406}},\ \bibinfo {pages} {486} (\bibinfo {year}
  {2000})}\BibitemShut {NoStop}%
\bibitem [{\citenamefont {Kokanovi\'{c}}\ \emph {et~al.}(2009)\citenamefont
  {Kokanovi\'{c}}, \citenamefont {Cooper},\ and\ \citenamefont
  {Matusiak}}]{Kokanovic2009}%
  \BibitemOpen
  \bibfield  {author} {\bibinfo {author} {\bibfnamefont {I.}~\bibnamefont
  {Kokanovi\'{c}}}, \bibinfo {author} {\bibfnamefont {J.}~\bibnamefont
  {Cooper}}, \ and\ \bibinfo {author} {\bibfnamefont {M.}~\bibnamefont
  {Matusiak}},\ }\href {\doibase 10.1103/PhysRevLett.102.187002} {\bibfield
  {journal} {\bibinfo  {journal} {Phys. Rev. Lett.}\ }\textbf {\bibinfo
  {volume} {102}},\ \bibinfo {pages} {187002} (\bibinfo {year}
  {2009})}\BibitemShut {NoStop}%
\end{thebibliography}
\rule[0.5ex]{0.8\columnwidth}{1pt}

\clearpage

%%%%%%%%%%%%%%%%%%%%%%%%%%%%%%%%%%%%%%%%%%%%%%%%%%%%%%%%%%%%%%%%%%%%%%
%%%%%%%%%%%%%%%%%%%%%%%%%%%%%%%%%%%%%%%%%%%%%%%%%%%%%%%%%%%%%%%%%%%%%%
%%%%% Supplementary Materials
%%%%%%%%%%%%%%%%%%%%%%%%%%%%%%%%%%%%%%%%%%%%%%%%%%%%%%%%%%%%%%%%%%%%%%
%%%%%%%%%%%%%%%%%%%%%%%%%%%%%%%%%%%%%%%%%%%%%%%%%%%%%%%%%%%%%%%%%%%%%%
% Section Formatting
\setcounter{secnumdepth}{1}
\setcounter{section}{0}
\setcounter{figure}{0}
\setcounter{table}{0}
\setcounter{equation}{0}

\renewcommand{\thesection}{S~\Roman{section}}
\renewcommand{\theequation}{S\arabic{equation}}
\renewcommand{\thetable}{S\arabic{table}}
\renewcommand{\thefigure}{S\arabic{figure}}
\renewcommand\theHfigure{S\arabic{figure}}

\onecolumngrid

\begin{center}
\large{\textbf{Supplementary Materials for\\[2ex]
Multiple Broken Symmetries in Striped La$_{2-x}$Ba$_{x}$CuO$_{4}$
detected by the \\ Field Symmetric Nernst Effect}}
\end{center}
\smallskip{}

\section{Sample Preparation\label{sec:SOM_MeasSetup}}

\begin{comment}
\textbf{S1.1. Sample Growth \& Preparation: }
\end{comment}
Single crystals of La$_{2-x}$Ba$_{x}$CuO$_{4}$ (near $x=1/8$)
were grown using the recently developed laser-diode-heated floating
zone method, which enables an exceptionally high degree of sample
homogeneity, as shown by the absence of impurity phases in Laue and
x-ray diffraction measurements\cite{Ito2013}. The samples were cut
along the crystal axes into rectangular bars for $a-b$ plane transport
measurements, with typical dimensions of $3.0\times0.5\times0.5$~mm
for in-plane resistivity ($\rho_{xx}$) measurements and $3.0\times1.5\times0.5$~mm
for thermoelectric and Hall measurements .

Electrical contacts were made with silver paste (DuPont\texttrademark{}
6838) baked in high purity O$_{2}$-flow environment at 450 $^{\circ}$C
for 10 minutes. The typical contact resistance achieved was $\sim0.5\,\Omega$.

\section{Magnetotransport Measurements\label{sec:SOM_Magnetization}}

\begin{comment}
\textbf{S1.1. Expt Conditions: }
\end{comment}
\begin{figure}[h]
\begin{centering}
\includegraphics[width=3.5in]{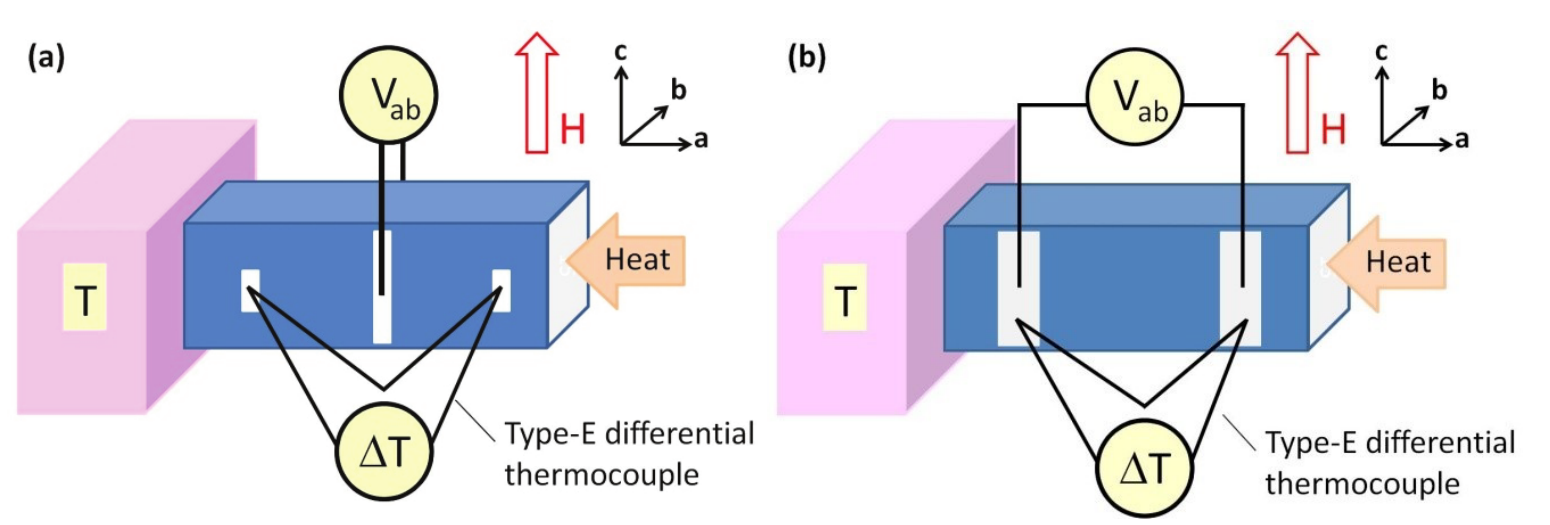}
\par\end{centering}
\protect\caption{\textbf{Thermoelectric Transport Schematics. }Schematic contact configurations
for thermoelectric measurements of the \textbf{(a)} Nernst and \textbf{(b)}
Seebeck coefficients. \label{fig:SOM_MeasSetup}}
\end{figure}
{} Thermoelectric Seebeck and Nernst effect measurements were performed
in homemade apparatus with low DC noise ($<\pm1$~nV), high temperature
stability ($<1$~mK), and under high vacuum conditions ($<1\times10^{-6}$~mbar).
Since thermoelectric measurements are in general sensitive to temperature
profile in the sample, heat exchange between the sample and its surroundings
has to be reduced in order to minimize the background temperature
gradient. Towards this, the sample was enclosed in a radiation shield
thermally anchored to the sample temperature, fine wires (25~$\mu$m
diameter) were used for electrical and thermal contacts, and the stage
temperature was carefully stabilized before data acquisition. Meanwhile,
supporting measurements of resistivity, Hall effect, and magnetization
were performed in Quantum Design\texttrademark{} Physical Property
Measurement System (PPMS) and Magnetic Property Measurement System
(MPMS) systems respectively, with AC transport measurements in current
bias mode at a frequency of 33~Hz.

\begin{comment}
\textbf{S1.2. Contacts \& Analysis: }
\end{comment}
\begin{figure}[h]
\begin{centering}
\includegraphics[width=5.8in]{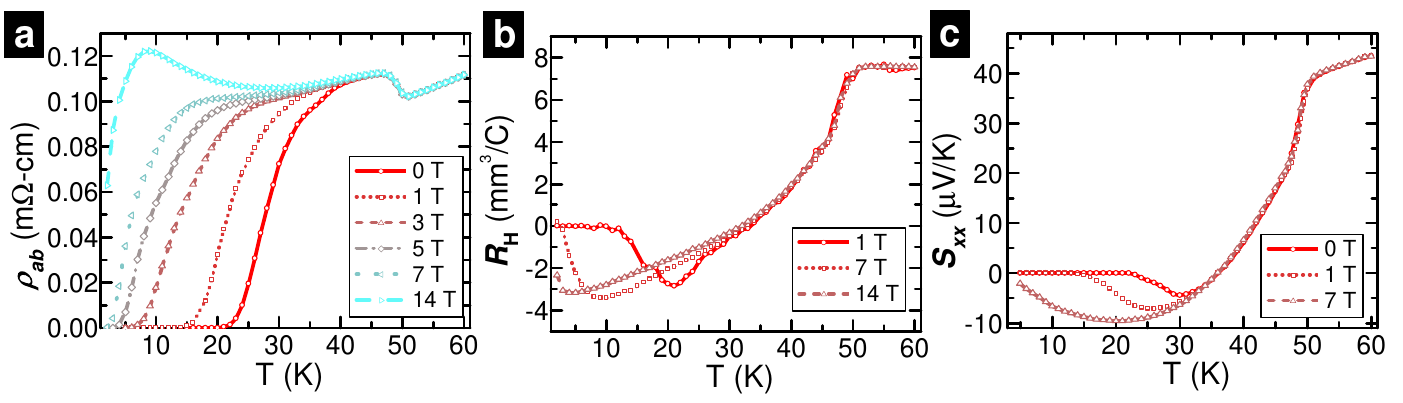}
\par\end{centering}
\protect\caption{\textbf{Conventional Magnetotransport Characterization. }Magnetotransport
characterization of the samples with doping $x=0.12$, showing \textbf{(a)}
the longitudinal resistivity $\rho_{ab}$, \textbf{(b)} the Hall coefficient
$R_{{\rm H}}$ and \textbf{(c)} the Seebeck coefficient $S_{xx}$.\textbf{
}\label{fig:SOM_MagTransport}}
\end{figure}
Schematics for the measurement of Nernst and Seebeck thermoelectric
coefficients are shown in \ref{fig:SOM_MeasSetup}. One end of the
sample is thermally anchored to the sample stage at temperature $T$.
Heat is applied to the other end through a 1 k$\Omega$ film heater
in contact with the sample. A typical heater power of $\sim1$~mW
generates a temperature gradient $\sim0.1$~K/mm, measured using
a pair of differential thermocouple wires (type-E, diameter 25~$\mu$m).
For Nernst effect measurement (\ref{fig:SOM_MeasSetup}a), voltage
contacts on the opposite sides of the sample were carefully aligned
in order to minimize longitudinal thermopower pickup. For the symmetric
Nernst effect results shown in this work (Fig 1-3), accuracy and reproducibility
of the results were ensured by measuring Nernst and Seebeck coefficients
simultaneously using the configuration shown in Fig. 2a.

\begin{comment}
\textbf{S1.3. Magnetotransport Data: }
\end{comment}
{} \ref{fig:SOM_MagTransport} shows the conventional magnetotransport
properties of our samples ($x=0.12$). The variation of these coefficients,
viz. the longitudinal resistivity $\rho_{xx}$, the Hall coefficient
$R_{{\rm H}}$ and the Seebeck coefficient $S_{xx}$, with magnetic
field is consistent with published literature\cite{Li2007c,Tranquada2008,Li2011d}.
Notably both Hall and thermpower data consistently show a sign change
at $\sim35$~K, emphasizing the self-consistency of these measurements.
Furthermore, the Hall data shows the absence of any hysteretic effects
that could support domain formation in the strip phase.

The temperature scales corresponding to the various stripe and superconducting
transitions that our samples undergo during these measurements (derived
from Fig. 2 and \ref{fig:SOM_MagTransport}) are listed in \ref{tab:TScales_Comparison}.

\begin{table}[h]
\begin{centering}
\begin{tabular}{|c|l|c|c|}
\hline
\multirow{2}{*}{\textbf{Temp. Scale}} & \multirow{2}{*}{\textbf{Significance}} & \textbf{Samp. \#1 (K)} & \textbf{Samp. \#2 (K)}\tabularnewline
 &  & \textbf{ ($x=0.12$) } & \textbf{($x=0.115$)}\tabularnewline
\hline
\hline
$T_{{\scriptscriptstyle LTT}}$ & LTT Transition & 50 & 48\tabularnewline
\hline
\multirow{1}{*}{$T_{{\scriptscriptstyle {\rm CO}}}$} & Charge Stripe Order & 47 & 45\tabularnewline
\hline
$T_{{\scriptscriptstyle {\rm SO}}}$ & Spin Stripe Order & \_ & \_\tabularnewline
\hline
$T_{{\scriptscriptstyle {\rm D}}}$ & Diamagnetism & 42 & 38\tabularnewline
\hline
$T_{{\scriptscriptstyle {\rm c,R}}}$ & Zero Resistivity & 22 & 25\tabularnewline
\hline
$T_{c}$ & Meissner State & 12 & 18\tabularnewline
\hline
\end{tabular}
\par\end{centering}

\protect\caption{Temperature scales of various transitions observed in the samples
studied in this work. \label{tab:TScales_Comparison}}
\end{table}

\section{Thermoelectric Measurements: Consistency Checks\label{sec:SOM_ThEl_Details}}

\begin{figure}[!h]
\begin{centering}
\includegraphics[width=2in]{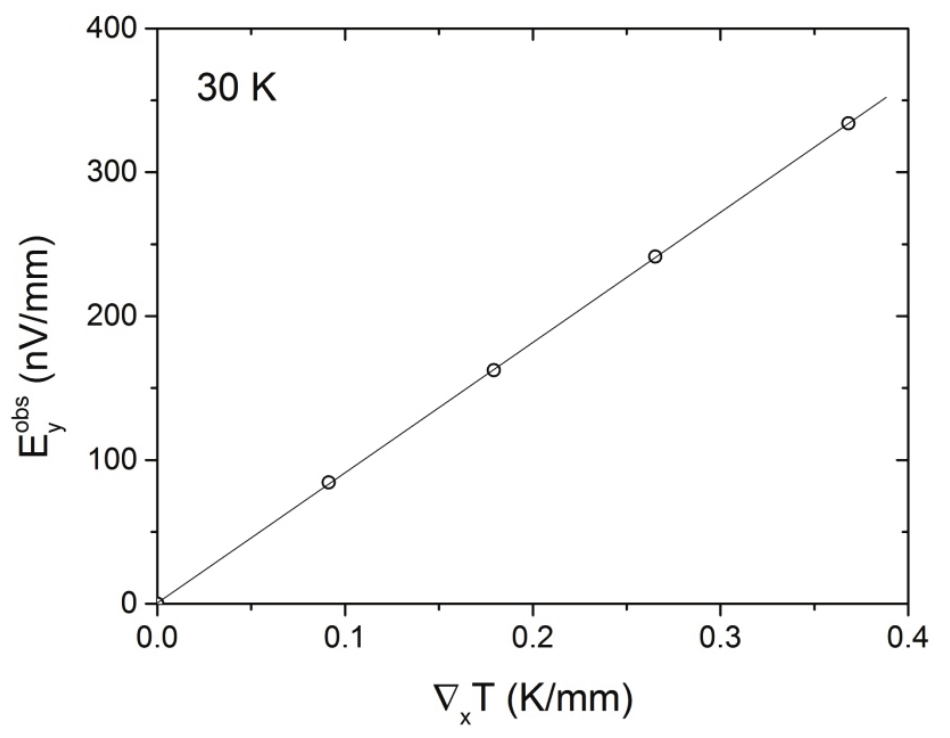}
\par\end{centering}
\protect\caption{\textbf{Linearity of Transverse Theroelectric Response. }Transverse
electric field response $E_{y}$ as a function of temperature gradient,
$\nabla_{x}T\sim0.1-0.4$~K/mm on $x=0.12$ at $T=30$~K. The response
is linear up to 0.4~K/mm. \label{fig:SOM_NernstHtrPow}}
\end{figure}

For correct determination of thermoelectric transport coefficients,
especially $S_{xy}$ (which is $\sim50\times$ smaller than $S_{xx}$),
it is crucial to ensure that the data is acquired in a linear response
regime, i.e. that the observed transverse electric field $E_{y}$
scales linearly with the applied temperature gradient $\nabla_{x}T$
for the conditions used in the measurements. \ref{fig:SOM_NernstHtrPow}
shows a representative selection of data acquired on LBCO ($x=0.12$,
$T=30$~K) towards Nernst effect measurements. With varying heater
power, resulting in temperature gradients $\nabla_{x}T\sim0.1-0.4$~K/mm,
the resulting transverse electric field $E_{y}$ is observed to be
linear throughout. This justifies the accuracy of the measurements
presented in this work, with typical $\nabla_{x}T\sim0.1$~K/mm --
well within the linear response regime.

\begin{figure}[h]
\begin{centering}
\includegraphics[width=4in]{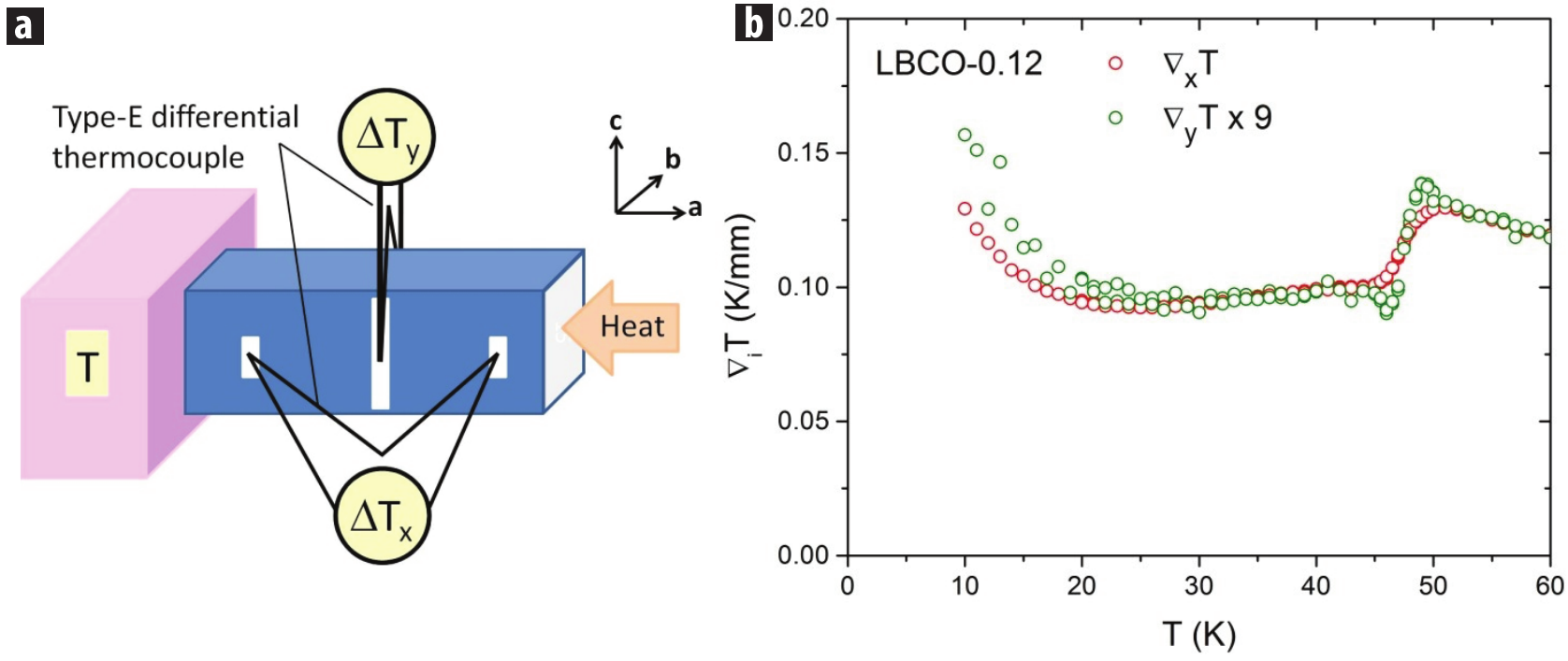}
\par\end{centering}
\protect\caption{\textbf{Thermal Gradient Measurements. (a)} Schematic contact configurations
for ``heat current deflection'' measurement. \textbf{(b)} Temperature
gradients along the $x$ and $y$ directions for $T<60$~K for $x=0.12$.
\label{fig:SOM_TempGradMeas}}
\end{figure}

The observation of a ZFN signal, can, in principle also result from
a spurious deflection of the longitudinal heat current towards the
transverse direction. Such a heat current deflection would result
in the generation of a Seebeck contribution along the $y$-direction,
manifesting as an artifact in the Nernst signal. To check for this,
we measured the temperature gradient along the $x$ and $y$-directions
$\nabla_{x}T$ and $\nabla_{y}T$ respectively, in response to a heat
current along the $x$-direction. This is accomplished by replacing
the pair of electrical contacts by another differential thermocouple,
as shown in \ref{fig:SOM_TempGradMeas}a.

\ref{fig:SOM_TempGradMeas}b shows that $\nabla_{y}T$ is non-zero;
however it results from a longitudinal $\nabla_{x}T$ pickup due to
the unavoidable misalignment in the thermocouple contacts (analogous
to Fig.~2(b)). Importantly, $\nabla_{y}T$ can be rescaled to overlay
on $\nabla_{x}T$ (rescale factor $\sim9$) for $T<60$~K. Any deviations
from the rescaling, which are near the limit of our measurement resolution,
can only result in spurious features an order of magnitude smaller
in size that the observed sharp feature at $T_{{\rm {\scriptscriptstyle CO}}}$.
Therefore, we do not observe a heat current deflection along the $y$-direction
considerable enough to explain the observed zero field Nernst effect
features in Fig. 2.

\end{document}